# Cosmology at the Crossroads of the Natural and Human Sciences:
## is demarcation possible?
### (A phenomenological insight)

Alexei V. Nesteruk

Department of Mathematics, University of Portsmouth,
Lion Gate Building, Portsmouth, PO1 3HU, Great Britain
alexei.nesteruk@port.ac.uk
http://userweb.port.ac.uk/~nesteruk/

## Abstract

The paper discusses the problem of demarcation between the dimensions of natural and the human sciences in contemporary cosmology. In spite of a common presumption that cosmology is a natural science, the specificity of its alleged subject matter, that is the universe as a whole, makes cosmology fundamentally different from other natural sciences. The reason is that in cosmology the subject of cosmological research and its "object" are in a certain sense inseparable. Any study of the universe involves two opposite perspectives which can be described as "a-cosmic" and "cosmic", egocentric and non-egocentric. Cosmology involves two languages, namely that of physical causality (pertaining to the natural sciences) and that of intentionality (pertaining to the human sciences). On the one hand the universe can be seen as a product of discursive reason, that is as an abstract "physical" entity unfolding in space and time. On the other hand the universe can be experienced through our participation in, or communion with the world understood as the natural context of living beings. This dichotomy between reason and experience, abstract construction and concrete participation, originates in the essence of human persons understood as unities of the corporeal and spiritual. On account of this dichotomy it is hard to set up a strict line of demarcation between the elements of the human and the natural sciences in cosmology. This confirms the intuition that any realistic world view is incomplete without a knowledge of what it means to exist as a human being. Conversely it is likewise impossible to understand human existence without considering its natural setting, that is the universe. We conclude that anthropology is incomplete without cosmology and *vice versa*.

## Preface

This paper represents a further development of ideas on epistemological meaning of modern cosmology which arose from research discussions between theology and cosmology and published previously in different resources. In particular, a variation of the first part of this text (pp. 2-22) has been published in Romanian under the title "Theological Commitment in Modern Cosmology and Demarcation between the Natural and Human Sciences in Knowledge of the Universe" ("Agajamentul Teologic în Cosmologia Modernă prin Problema Demarcației între Științele Naturale și cele Unabe în Cunoașterea Universului" Sinapsa (Editura Platytera, Bucharest), Nr VI, 2010, pp. 25-44. Some ideas on application of the idea of coherence of epistemic justification in cosmology supporting our claim on the strong presence of the human sciences' elements in cosmology were published in A. Nesteruk, *The Universe as Communion*, London, T&T Clark, pp. 244-46; "From the Unknowability of the Universe to the Teleology of Reason", In J. Bowker (Ed.) *Knowing the Unknowable: Science and Religion on God and the Universe*, I. B. Tauris & Co Ltd, 2009, pp. 63-86. (Amended Russian translation: "From the Unknowability of the Universe to the Teleology of the Human Spirit: a Phenomenological Analysis in Apophatic Cosmology", In A. Grib (Ed.) Scientific and Theological Reflection on Ultimate Questions: Cosmology, Creation, Eschatology, Moscow, St. Andrew's Biblical and Theological Institute, 2009, pp. 101-21.) A suggestion to treat the universe as a "saturated phenomenon" (which obviously deviates from a naturalistic approach to the universe) was made in our paper "Transcendence-in-Immanence in Theology and Cosmology: a New Phenomenological Turn in the Debate", *Studies in Science and Theology* (Yearbook of the European Society for the Study of Science and Theology 2009-2010), vol. 12, 2010, pp. 179-198. (Amended Russian translation: " 'Transcendence-in-Immanence': a new phenomenological turn in the dialogue between theology and cosmology", In V. Porus (Ed.) *Scientific and Theological Epistemological Paradigms*, Moscow, St. Andrew's Biblical and Theological Institute, 2009, pp. 35-62).

**Introduction**

Contemporary physical cosmology is a well established and vast enterprise which includes astronomical observations, space programmes, research institutions and funding strategies. Cosmology develops fast. Every day one discovers dozens of new publications on the internet archives. Monographs and popular books telling stories about the universe, about its study and those who study it are in abundance. Cosmological ideas are used and misused in science fiction and fiction in general. Cosmology becomes a sort of a cult reading as if humanity touches upon something ultimately sacred and indispensable for its life. Cosmology gathers numerous conferences, workshops and public lectures resulting in further publications of collective volumes. Apart from physical scientists, cosmology attracts historians and philosophers of science, as well as millions of those who adore science and trust its final word on the nature of things. This is a dynamic set of enquiries about the world around us which constitutes an integral part of contemporary intellectual culture. It is exactly the popularity of cosmology in mass-media and among ordinary people which manifests that it affects a collective consciousness of people and has existential and ethical implications. Cosmology is involved in the dialogue with religion: it becomes an arena of theistic inferences and justifications of otherworldly transcendence when the results of its theories are brought into correlation with theological convictions. Contemporary cosmologists are often seen as exercising a certain priestly role in modern society as if cosmological ideas had an immediate existential and social impact that would catch and fascinate public opinion.

In spite of all these facts  cosmologists' confidence in cosmology's ability to explain the essence and contingent facticity  of the universe is far from being justified. Some cosmologists raise doubts whether cosmology can pretend  to be following rigorously what is called scientific method[1], understood so that all knowledge including mathematical theories leads to experimental verification.  It is a fact of cosmology's sociology that there are extrapolations and conjectures in cosmology's claims for truth which go beyond scientific justification and this involves the whole field into an interdisciplinary discourse in which the criteria of validity and truth are much more vague that those in the natural sciences. Correspondingly, the objective of this paper is to elucidate the nature of cosmology's claims for the value and truth of its theories in a cross-disciplinary context related to knowability of the universe and its relation to human agency, its history and self-understanding. This will be done in the perspective where cosmology is treated as a mode of human activity contributing to the "infinite tasks" of humanity, its culture and spiritual advance.

---

[1] A careful analysis of methodological weaknesses in cosmology has been done in a paper of George

**Physical cosmology and an input of philosophy**

Cosmology, understood as part of theoretical physics, forms a subject matter that by its very nature tests the boundaries and the very possibility of scientific explanation. Indeed, cosmology describes itself as a science which deals with "the universe as a whole", the universe as the all-encompassing, singular and unique "object" of cosmology. However the usage of the word "object" applied to the universe as a whole is problematic simply because the mainstream understanding of objectivity does not allow the concept of the universe to fit in it. Indeed, according to this view the universe consists of independent individual things (objects) which are embedded in space-time. These things as objects are *individuals*, because they have a spatio-temporal location, they are a subject of predication of properties, and they are distinguishable from each other through some properties. The universe as a whole cannot be thought as an object (or as an individual) because it is as a whole not embedded in space-time (it is rather a totality of space and time which transcends their characteristic features such as extension). [2] The universe is unique and cannot be distinguishable from anything through particular properties because, by definition, it comprises everything. The predication of the universe as a whole in terms of properties is problematic because the universe does not attain original givenness in the manner characteristic of particular individual things.[3] The constitution of an individual thing as an object, that is as a thing subjected to thematisation and objectification assumes as a condition the release from "environmental confinement"[4] or the context in which a thing is looked at. The universe as a whole cannot be released from such a confinement because, in a way, it is itself, by definition, the ultimate environment and context for everything.[5] Thus the standard meaning of the phrase "object of explanation" as if its identity has been defined cannot be applied to the universe with any ontological clarity.[6] But we do indeed understand and use this expression "the universe" and therefore there must be a manner in which the universe is given, a consciousness of the universe that bestows sense on such language. This implies that before any philosophical deliberation or scientific thematisation of the universe, there must be experience of the universe as the recognition that there is the permanent and persistent in the background of change or the variable. There is the sense of identity of the universe as an intentional correlate of subjectivity, but the identity as ideal and unfulfilled. A possible scientifically reductive approach to identity of the universe as an inherent and non-relational aspect of an entity or a logical subject does not clarify the ontological status of this identity. Within these, so to speak, existential delimiters, the universe of cosmology, being thematised, naturally represents the ultimate noematic limit in the process of scientific exploration and explanation. Nothing further is empirically or theoretically accessible to which recourse can be made in order to explain the most general properties of the universe as a whole and the facticity of its own existence.[7] In one way or another, natural scientific explanation stops right there.

---

[2] The word "object" cannot be legitimately applied to the universe because the universe as a singular and self-contained whole cannot be detached from human insight and thus positioned as something which is outside and devoid of the human presence. There is a fundamental inseparability between the universe and the knowing subject who is always a part of the universe. Another problem emerges from the universe's uniqueness which cannot be set among other universes. The modern view of the universe as developed from a singular state (Big Bang) which *de facto* encodes the universe in its totality invokes a counter-intuitive sense of the universe as a singular, that is unrepeatable *event* (not an object!) with respect to which the natural sciences experience the sense of fear and desire to explain it away. Edgard Morin expressed the idea that the cosmos reveals itself as the universe and event. On the one hand the physical universe constituted through regular repetitive features, on the other it is a singular event as phenomenon, the phenomenon which evolves for more than ten billion years. The temporal unfolding of the universe which appears to human contemplation lies indissolubly in the advent-event (avènement-événement) of the world (E. Morin, *Science avec conscience*, Paris, Fayard, 1982, p. 120).

---

[3] It is because of the inseparability between the human observer and the universe that the conditions of the universe's observability and mathematical expressibility are constitutive of the very concept of the universe. In this sense the "physical objectivity" of the universe cannot bear an independent reality in a classical sense. Indeed, unlike in classical physics, the basic conditions of the constitution of the universe as a whole have not been permanently available and thus have to be questioned (Cf. Bitbol, M., Kerszberg, P., Petitot, J. (Eds.) *Constituting Objectivity. Transcendental Perspectives on Modern Physics*. Springer, 2009, pp. 4, 18.)

[4] Terminology of M. Heidegger (See *Being and Time*, Oxford, Blackwell, 1998, p. 413).

[5] Here one sees an original sign that cosmology in a way has some features of the human sciences, because it is known that the release from environmental confinement is not necessary for thematisation and objectification in the human sciences, where a perspective on reality is crucially dependent on the researcher's intentionality originating in the existential and socio-historical condition (and thus cannot be environmentally free). Applied to cosmology this would mean that if one implies that the in-itself of the universe (as its identity) be studied, it must preserve this identity as free from any change through the release from environmental confinement, that is from the inherent subjectivity of a knower of the universe.

[6] C.f. Theses A1 and A2 in G. Ellis, *Issues in the Philosophical Cosmology*, p. 1216.

The very existence of the universe turns out to be the precondition for physical science: the latter describes and explains phenomena which take place in the universe as something which is already given. This is the reason why the universe (as the totality of being) is not itself subject to a physical explanation. The phenomena with which physics deals have to be present. Physics simply takes the existence of its objects for granted. The laws of physics are laws that hold *within* this universe; they do not purport to be laws that hold across "universes" (which in this sense would be universal for many universes with different contingent properties), whatever that would mean. Physics is not able to enquire into the underlying facticity of the phenomena within the universe. If this facticity is associated with the contingent appearance of phenomena (as contingent outcomes of physical laws[8]) as if these phenomena manifest the radical coming into being of that which has not been before, physics definitely cannot link the being of these phenomena with that something (non-being) they come from. In other words, physics can deal with the manifestations of being but not with the ground of these manifestations in "non-being". It deals with something that obeys laws which are already in being. In technical philosophical language the same idea can be expressed differently: since physical cosmology is capable of apprehending the interior of the universe, the universe exhibits itself as intelligible; but because of the contingent nature of this intelligibility (it cannot explain itself, otherwise it would not be contingent) the universe embodies a semantic reference beyond itself. Cosmologists cannot themselves deal with this "other-worldly" reference and conduct a proper philosophical work. Some physicists in an attempt to address the foundational questions in cosmology make manifest a "philosophical" mode, not because they adhere to a realm of "philosophy" but because they do not follow the normal ways of theory-assessment in the natural sciences. This was the original motivation, for example, for inflationary cosmologies which aspired to explain away the problem of the special initial conditions of the universe responsible for the contingent display of the astronomical universe. A similar motivation lies in ideas of multiverse. However, these models, having a developed mathematical basis and being employed for problem-solving, raise philosophical problems and need competence and appraisal through borrowing methods of philosophy and insights of the human sciences.

---

[7] This was always realised by cosmologists themselves. As an example one can refer to D. Sciama's interview of 1978 where he underlined the existence of a borderline between the ultimate questions about the universe's facticity and the exploration of its properties: "None of us can understand why there is a Universe at all, why anything should exist; that's the ultimate question. But while we cannot answer this question, we can at least make progress with the next simpler one, of what the Universe as a whole is like." (Quoted in H. Kragh, *Cosmology and Controversy. The Historical Development of Two Theories of the Universe*, Princeton University Press, 1996, p. xi).

[8] However the very contingent appearance of things in the universe points towards the laws whose outcomes supply these appearances: there must be these laws in order to have these particular things. It is difficult to separate in the universe as a whole between its factual (material) and nomic (law-like) features. In this sense one can talk about facticity of physical laws themselves as linked to the boundary or initial conditions in the universe. See, for example, in this respect Yu. Balashov, "On the Evolution of Natural Laws", *The British Journal for the Philosophy of Science*, vol. 43, No. 3, 1992, pp. 343-370 (354-56); "Two Theories of the Universe. Essay Review." *Studies in History and Philosophy of Modern Physics*, vol. 29, No. 1, 1998, pp. 141-149 (147); "A Cognizable Universe: Transcendental Arguments in Physical Cosmology", in M. Bitbol, P. Kerszberg, J. Petitot (Eds.) *Constituting Objectivity. Transcendental Perspectives on Modern Physics*. Springer, 2009, pp. 269-277.

One can generalise by saying that on the one hand physical cosmology avoids touching upon ultimate questions; on the other hand, because of the special status of its subject matter, that is the universe as a whole, as well as the fundamental inseparability of human subjectivity from the universe, cosmology is imbued with these questions and in order to attend to them one has to invoke a philosophical attitude to cosmology.[9] By conducting a philosophical analysis of cosmology one can on the one hand articulate the *qualities* of cosmological theory which make it scientific, and identify the naturalistic *limits* within cosmological methodology. On the other hand, by transcending these limits through an enquiry in cosmology's facticity, one inevitably brings cosmology beyond the scope of the natural sciences since, de facto, here humanity enquires into the facticity of its own historically contingent subjectivity. Philosophy here manifests itself as a method of enquiry into the sense-forming activities of human subjectivity in the subject area of the universe as a whole.[10]

---

[9] It is this mentioned inseparability which makes the cosmological idea (that is the idea of totality of the world) fundamentally different among other ideas of reason, such as the idea of soul or the idea of God. Kant wrote that neither psychological nor theological idea entail contradiction and contain antinomies. (I. Kant, *Critique of Pure Reason*, (A673/B701). 2nd ed. Trans. N. K. Smith. London: Macmillan, 1933, p. 552). Practically this means that one can easily deny the existence of a soul (let us say, on materialistic grounds) or deny the existence of God (on atheistic grounds). However it is impossible to deny the existence of the universe for it would deny the empirical world of sense which is part of the universe and which contains the foundation of all knowledge about universe. The antinomian nature of reasoning about the universe originates exactly here: by being in the sensible world one cannot disentangle from the universe, at the same time the universe as totality is never fully materialised in the world of the senses.

[10] The fact that the encounter with the problem of the universe as a whole represents more an epistemological issue than anything which can be associated with the natural sciences, was understood long before by such thinkers as Nicholas of Cusa and Kant. The very concept of "learned ignorance",

However, since philosophers do not have a supply of knowledge about nature in advance, on which they can draw or to which they should refer, it would be wrong to take the philosophy of cosmology as dealing with issues independently of the research going on in physics and mathematics. But in spite of the obviousness of the fact that the origin of scientifically motivated facts lies within cosmologists' thinking, the sense of cosmological ideas and their significance for the constitution of the noetic pole of the enquiry (that is for human subjectivity), exceeds the scope of the natural sciences and thus requires an appeal to the methods of those sciences which are not restricted in their scope to the causality of nature.

**The special status of cosmology as a natural science: from substance to manifestation**

The special status of cosmology among natural sciences is determined by the decisive factor that its subject matter is unique and cannot be represented as an outside object, so that there is a fundamental inseparability of the enquiring intellect and the universe as a whole. Said philosophically, the universe enters all forms of human cognition as the ultimate horizon of contexts.[11] Here we are confronted with a question about the status of cosmology as a natural science. In an attempt to study *some particular aspects* of these contexts cosmology exhibits certain features of the human sciences in the sense that the humanly made choice and emphasis of topics of investigation which amounts in modern terms to the apophaticism of knowledge in general, and which had been drawn from astronomical-cosmological considerations, had most of all an epistemological meaning pointing toward the limits of reason and puzzles which it has to encounter while dealing with such a limiting concept as the universe. (See, for example, A. Koyré, *From the Closed World to the Infinite Universe*. New York, Harper & Brothers Publishers, 1958, pp. 5-19). A similar sense was attached by Kant to his famous cosmological antinomies, which were indications of the fundamental paradoxical structures of reason rather than any constructive theories of the universe. Here is a characteristic quote from a contemporary treatise on Kant: "Because reason examines itself in order to extract laws from within itself, instead of simply greeting these laws, the cosmological antinomy is the place where the innermost depths of our humanity manifest themselves. In the antinomy, nature speaks to our inquiring minds in the most direct possible way, precisely because, as a complete whole, it is exposed to the danger of being lost in obstinacy or despair."(P. Kerszberg, *Critique and Totality*, State University of New York Press, 1997, p. 101).

[11] Here, in analogy in the Husserl's definition of the "world-horizon" the universe as such is never given in a manner pertaining to ordinary objects. The universe as a horizon of all contexts in the physical and mathematical enquiry in the structure of the world cannot be an object and is distinct from any object given in the background of contexts. The universe is coperceived as the necessary horizon of all individual beings (astronomical or terrestrial) which are immediately experienced. (E. Husserl, *Phenomenological Psychology*. The Hague, Nijhoff, 1977, pp. 70-73; see also A. Steinbock, *Home and Beyond. Generative Phenomenology after Husserl*. Evanston, Northwestern University Press, 1995, p. 104.)

through their naming, methods and goals have a genetic *historical* priority over the post-factum made non-egocentric claims about the reality of the universe as if it is in itself. The same is true with respect to any part of physics. However, the seeming epistemic priority of the human element in cosmology is linked to the fact that the human world (or the "premise-world") associated with the conditions of embodiment has object-noematic priority over all "other worlds", for cosmology, unlike other sciences, has to predicate that reality which is far away in a generic sense from the premise-world. This predication is being made not only as a bottom-up explanation (that is based on ascending series of physical *causation* from the macroscopic empirical phenomena to the additive totality), but also as a top-down *inference* based on the workings of the *intentionality* of human subjectivity.[12] This intentionality includes, for example, the very idea of the universe as the overall totality. From the point of view of empirical physics the invocation of this idea is optional. Correspondingly the idea of the origin of the universe does not proceed from earthly physics: it enters the discourse through an intentional interrogation into the ground of the universe's facticity, an interrogation which is not part of the enquiry into physically causality, but rather is a philosophical quest for the sense of being. To understand, in an existential sense, intentionality invokes intelligible, invisible entities in order to "explain", or more precisely, to interpret the phenomenal. Cosmology in this respect provides an endless chain of illustrations.

When we accentuate the presence of the language of intentionality in cosmological discourse we effectively involve physical cosmology (which is by its status a science of the abstract, and detached from human reality, universe) in the context of the human affairs, thus exhibiting in a characteristic way that intrinsic ambivalence in cosmology which originates in the paradoxical human condition as

---

[12] This distinction can be elucidated by a quote from a paper of C. Harvey: "It is common parlance to say that whereas the natural scientists seek to explain, the human scientists seek to understand. This distinction between understanding and explanation, however is itself predicated upon the deeper distinction between *intentionality* and *causality*. If the natural sciences rely upon physicalistic causality as the human sciences rely upon intentionalistic motivation, and the intentionalistic motivation is shown to be prior to causal rationality, then natural science will be shown to be posterior to, because ultimately explainable in terms of, human scientific motifs". "Natural Science is Human Science. Human Science is Natural Science: Never the Twain Shall Meet". (In B. E. Babich, D. B. Bergoffen, S. V. Glynn (eds.) *Continental and Postmodern Perspectives in the Philosophy of Science.* Aldershot: Avebury, 1995, pp. 121-136(125), (emphasis added).

"both being a subject of the world and being an object in the world".[13] Being a subject of the world man articulates the whole universe on the grounds of its existential inference of its commensurability with the universe. Being an object, human being realises its insignificance for the whole universe and thus its incommensurability with it. It is the sense of commensurability which is embedded in cosmologist's *intentionality* of believing in and predicating of the universe as a whole. And it is the sense of incommensurability which is implied by cosmologists' physical embodiment that advances their search for the structure of the universe based on physical causality. In spite of its paradoxical standing this twofold perception of the interplay between humanity and the universe reflects an inevitable feature of any disclosure of being by human agency. In this sense the unity of opposites in this paradox is still preserved by the uniqueness of humanity as the centre of disclosure. Correspondingly any pretence of sheer objectivity for the knowledge of the universe as a whole is blatantly incorrect so that a simple relief from this tension would be to conjecture that the *content* of cosmological knowledge (that is, astronomical facts and theories of the universe as a whole including its alleged origins) should be considered not as contraposed and "transcendent" to human subjectivity, but as *transcendentally constituted*. In other words cosmology itself must be seen as part of the transcendental discourse, that is the discourse of the conditions which allow the universe to manifest itself (in particular, through mathematical expressibility). Correspondingly one should make a subtle distinction between the principles which coordinate knowledge of the universe and those connecting principles (expressed mathematically) which state the relation between the properties of objects which are already constituted. It is this transcendental constitution which, being restricted by the outer universe through the stabilisation of patterns of thought, has a fundamental human origin in the very act of its intentional launching, that is an expression of interest and participation in that which gives itself for being constituted.

---

[13] This paradox is a perennial problem of philosophy and was anticipated by ancient Greek philosophers and Christian thinkers. It was express differently by such philosophers as Kant (see, for example, Kant's conclusion to *Critique of Practical Reason*.) Among phenomenological philosophers who dealt with this paradox one can mention E. Husserl, M. Scheler, M. Merleau-Ponty, E. Fromm and others. The general discussion of this paradox can be found in D. Carr, *Paradox of Subjectivity*, Oxford University Press, 1999. The decisive role of this paradox in discussion on science and theology can be found in A. Nesteruk, *The Universe as Communion*. London: T&T Clark, 2008, pp. 173-175). Applied to the study of the universe the paradox of human subjectivity can be formulated as follows: on the one hand human beings in the facticity of their embodied condition form the centre of disclosure and manifestation of the universe as a whole, must be rather considered it as overall-space and time which exceeds the limits of the attuned space related to humanity's comportment on the planet earth (the home place). On the other hand the depicted universe as a vast continuum of space and time positions humanity in an insignificant place in the whole totality making its existence not only contingent (in physical terms) but full of nonsense from the point of view of actually infinite universe. Said bluntly the actual infinity of the universe is attempted to be articulated from an infinitely small part of its formation. One could express this differently: through its insight humanity is co-present in all points of what it observes in the universe, or imagines while physically being restricted to an insignificant part of it. Cosmology as the discourse of the universe as a whole brings one face to face to a general philosophical objective of avoiding any sort of foundationalism in knowledge of the universe which insists on the ground-grounded relation between humanity and the universe leading either to an idealistic reduction (subjectivity as the ground of the world) or to a materialistic, mathematically deterministic diminution of consciousness to illusion. In either mode of reduction the reality of the ground absorbs the grounded and the grounded is reduced to the categories of the ground. To avoid these reductions, the embodiment, as a premise of the person's grasp of the world, must be taken considered as that "over here", where a particular and immediate indwelling of life and the universe comes to *presence*. It is this coming to presence that determines that "place" which constitutes person as a centre of disclosure and manifestation of the universe.

Seen in such a way the intended "subject matter" of cosmology (the universe in its totality) exceeds the scope of the physical sciences for it refers not only to the content of what has already been manifested, but to the conditions of this manifestation which are not part of the physical description *per se*. Seen in this perspective only, the phenomenal universe is a sort of a static image in the ongoing process of manifestation. By its constitution, physical cosmology provides us with a particular, logically and physically accessible pattern in the interpretation of the universe which, however, does not exhaust the whole sense of human presence in the universe as the condition for its manifestation.[14] The transcendental sense of cosmological discourse arrives from the recognition that the universe is not that which is manifest, but that it is *the manifestation* related to humanity. In this sense the universe is always our universe. By its sense the discourse of the universe as *the manifestation* has to comprise not only the current scope of observations and theories about the universe, but the whole history of formation of views on the cosmos as well as all philosophical and theological issues on the conditions of knowledge of the universe, the *telos* of this knowledge and its value. The universe as manifestation implies a constant participation or communion with it which is tantamount to saying that the universe as manifestation means life.

---

[14] This concerns first of all the dimension of personal (hypostatic) embodiment. Indeed the discursive or linguistic expression of experience of the universe does not rule out the immediate corporeal presence of the universe on the level of sheer consubstantiality between human beings and the universe. Correspondingly if this dimension is overlooked then the perceived inability of cosmology to make results personally meaningful can be alienating and frustrating for non-specialists: for example, the sheer insignificance of humanity on the cosmic scale can create a sense of anxiety and despair related to the meaning of human life. However cosmic physics does not exhaust the sense of the human experience of space, or astronomical objects. Our experience of the universe as that mysterious environment with beautiful night skies and warming presence of the life-giving sun exceeds and is much richer than just knowledge of astronomy or solar physics. The problem is that the formalised and mathematised science sometimes has the effect of de-legitimising and de-appreciating other ways of communion with the wonders of space. (Cf., e.g., A. Nieman, "*Welcome to the Neighbourhood*: Belonging to the Universe", *Leonardo*, vol. 38, No. 5, 2005, pp. 383-388).

The conditions of manifestation of the universe which are always implicitly present behind its empirical appearances and theoretical representations yet escape an explicit constitution. They reveal themselves through an excess of intuition over logical simplicity and mathematical thoroughness which delivers the paradoxical sense of *presence* of the universe, the sense which is never disclosed in discursive terms thus leaving one with an immanent awareness of the universe's *absence*. Put differently, the universe *is*, but there is no answer to the questions "What is it?" The incompleteness of any physical description of the universe brings us to that stance in knowledge which is called "apophaticism", that is a mode of experience in which that which is intended to be signified through discursive description is never exhausted through its signifiers.[15] The ambiguity of the presence in absence of the universe deprives a genuine cosmological project of any flavour of *foundationalism* understood as an epistemological correlate of the notion of an ontological ground be it the constituting subjectivity of the self, or the outer universe as underlying substance. Cosmology has to function in the conditions of the classical paradox of human subjectivity in the world which arises in this context and points to the fundamental difficulty in attempting to formulate the ontology of the universe in terms of ground-grounded relationship. The universe as manifestation thus escapes any accomplished definitions and descriptions and, because of this, human subjectivity itself is being constituted through its openness to the universe to the extent it cannot comprehend the universe. One sees thus that cosmological discourse (as a mode of the natural sciences) cannot pretend to be complete without recourse to the essence of the agency disclosing the sense of the universe, that of human beings. [16]

---

[15]One can mention that the "apophatic" conviction applied to some limiting situations in cognition is well known in history of philosophical and theological thought. Generalising this conviction towards knowledge in general, C. Yannaras describes "as "apophatic" that linguistic semantics and attitude to cognition which refuses to exhaust the content of knowledge in its formulation, which refuses to exhaust the reality of things signified in the logic of signifiers (C. Yannaras, *Postmodern Metaphysics*, Brookline, MS: Holy Cross Orthodox Press, 2004, p. 84). In philosophy, for example, it originates from an epistemological argument pertaining to a sort of linguistic reformulation of the Kantian transcendentalism (which is typical for post-structuralism) that language conditions the accessibility and intelligibility of reality. In this approach the very phrase "there is" points to a referent which the very language cannot capture because the referent is not constituted by language and by definition is not the same as it linguistic effect. According to this view there is no access to the referent outside the linguistic effect, but the linguistic effect is no the same as that referent it attempts but fails to capture. This situation entails, in analogy with theology, a variety of ways of making such a reference, where none of which can claim it exclusiveness and true accessibility to what the reference is made. A phenomenological philosopher J. Ladrière, without using the notion of apophaticism, points towards the same feature of any knowledge, more precisely to the apophaticism of that fashion in which the human existent approaches the encounter with the world. An object is never a pure reference to itself, but is also a revelation of the fashion of its comprehension. (J. Ladrière, "Mathematics in a Philosophy of the Sciences", In T. J. Kiesel and J. Kockelmans (eds.), *Phenomenology and the Natural Sciences*. Evanston: Northwestern University Press, 1970, pp. 443-465 (448, see also p. 450)). The range of cognitive situations which fall under the scope of apophaticism can be found in works of J.-L. Marion under the name of "saturated phenomenon" (see his *In Excess. Studies of Saturated Phenomena*. New York, Fordham University Press, 2002).

**The nature of manifestation and ontological commitment**

In some cases cosmology claims the existence of things on the grounds of theoretical consistency and a fit with other plausible constructs, but for which we can have no observational evidence (that is, the principle of direct correspondence with empirical reality is not applicable).[17] Such a situation, for example, happens in the extreme case of the construct of the multiverse[18], where no direct observational or experimental tests of the hypothesis are possible, and the assumed underlying physics is probably untestable in principle. These possibilities do not by themselves prove correct epistemic justification, even less do they point to the truth-content of what theories claim. It is seen that here a sort of philosophical, that is trans-scientific insight is invoked.

In the case where cosmology predicates things beyond their verification through correspondence it appeals first of all to the method of extrapolation (understood in a wide sense) which itself must be evaluated as tacitly committed to a sort of realism grounded in belief of the efficacy of extrapolation. Philosophically and scientifically the problem of extrapolation arises from those limits of scientific explanation which are set by the observational constraints inherent in our earthbound home-place. All that is in principle directly accessible to observations is positioned on the surface of the past light-cone with its apex on the planet Earth.[19] Outside that cone one has the uncertainties of extrapolation.[20] Thus the extension of a cosmologist's insight into the universe from earth, in the attempt to encompass the universe in a single vision (including its absolute origin), requires an inference from what is already known to what is as yet only conjectured. For a form of knowledge that rests its claim on its empirical, observation-based, access to the world (most of the natural sciences), these limits raise clear difficulties.

---

[16] C.f. "A philosophy of nature and a philosophy of man are mutually complementary;… neither can be completed unless it shows itself as the counterpart of the other", G. De Laguna, *On Existence and the Human World*, New Haven and London, Yale University Press, 1966, pp. 81-82.

[17] This, for example, can be related to the cosmological principle which postulates uniformity of the universe beyond observational limits. Another example is a famous "inflaton" field which drives the exponential expansion of the early universe.

[18] Multiverse proposals in cosmology refer effectively to the old idea of the plurality of worlds understood either in a physical sense as an ensemble of worlds with all possible physical conditions, or a variety of mathematical structures which have or do not have their incarnation in the physical. In this case the existence of our universe in its contingent facticity is explained away through a reference that it simply belongs (in a generic sense) to an ensemble of universes which through its totality contains whatever is possible. (The literature on the multiverse is vast, as an example see a paper M. Tegmark, ("Parallel universes". In *Science and Ultimate Reality: From Quantum to Cosmos*, Eds. J. D. Barrow, P. C. W. Davies and C. Harper, Cambridge University Press, 2003, pp. 459-491) or a book edited by B. J. Carr (*Universe or Multiverse*, Cambridge University Press, 2007) with a variety of papers on different aspects of the multiverse debate.) In all multiverse proposals the question of existence, that is of the contingent facticity of this universe, is thus quite illegitimately transferred to the question of selection, whereas the issue of the existence of the multiverse itself cannot not addressed at all for obvious philosophical reasons.

One could claim that "extrapolations" (inferences) towards the fundamentally non-observable and untestable are simply physical hypotheses that are assessed along a variety of lines including observational tests as one of them. These hypotheses may rely on appeal to analogy, on consistency with other cosmological contexts, on logical fertility and explanatory force, or a mathematical consistency and elegance. Over time they may be woven into a more and more tightly connected set of beliefs and ideas, each element of which derives support from the set as a whole.[21] One can claim even further that *extrapolations* in cosmology itself (whatever this means, including a shift of "home places"[22] in the cosmological principle, or a free eidetic variation[23] of the parameters of the whole world which happens in theories of multiverse) implies an extended sense of "scientific justification", for example epistemic coherence which does not necessarily refer to tests and observations. This, in turn, entails a different commitment to realism.

---

[19] There is a tiny piece of the human observer's world line which relates to the immediate cosmic environment like the earth, planets in the solar system, stars in our galaxy which, in terms of cosmic times and thus space, are "close" to us so that their separation from us is in a way "commensurable" with the humankind's life span. We assert the existence of such objects in terms similar to those of the earthly objects.

[20] Thesis B1 in Ellis' "Issues in the Philosophy of Cosmology", p. 1220.

[21] See, for example, E. McMullin, "Long Ago and Far Away: Cosmology and Extrapolation". In R. Fuller (ed.), *Bang: The Evolving Cosmos*, Saint Peter, Minnesota: Gustavus Adolphus College, 1994, pp. 119-120.

[22] This is the terminology of E. Husserl; see his paper "Foundational Investigations of the Phenomenological Origin of the Spatiality of Nature" In P. McCormick, F. A. Elliston (Eds) *Husserl Shorter Works*, Indiana, University of Notre Dame Press, 1981, pp. 222-233.

[23] On eidetic variation in phenomenology see e.g. R. Sokolowski, *Introduction to Phenomenology*, Cambridge University Press, pp. 177-184.

For example, in the models of origin of the universe, the major presumption is that one can extend the laws of physics (comprehended by us through mathematical formulae) towards something which can not be physically independent of its mathematical gestalt. In other words, such an extension presumes effectively a set of beliefs is possible that catch the sense of reality beyond the sensible (corporeal, in a sense of physical equipment as extension one's bodily function) as its efficacious identity (which could be either on the level of the alleged substance or on the level of ideal forms) through time in spite of the postfactum resistance of reality to this.[24] The validity of these beliefs can only be justified on the grounds of their coherence as well as, to a lesser extent, agreement with that border-line physics which through observation is linked to the empirical validation. The situation when justification is linked to beliefs is dealt with by that part of contemporary epistemology which is called the coherence theory of epistemic justification and which holds that a belief is justified to the extent to which the belief-set of which it is a member is coherent[25]; what is at issue in a coherence theory is a matter of a proposition's relation to other propositions, and not its 'coherence' with reality or with the facts of matter.

Now we see that it becomes a task for philosophy to discuss the various sorts of hypothetical extrapolation that cosmologists make as a regular part of their work and the implied philosophical beliefs which drive them. As a matter of illustration let us refer to the basic assumption underlying the very possibility and foundation of modern cosmology, that is the principle of uniformity of space-time and matter (cosmological principle) which is based in extrapolation (in the certainty of a belief in an indifferent location of humanity in the universe) that the average isotropic picture of the large-scale distribution of matter in the universe as observed from the Earth can be transferred to all possible locations (thus implying spatial homogeneity).[26] This extrapolation makes manifest a certain *philosophical* and, may be, even a *theological commitment* which acts in the cosmologist's mind as a regulative and indemonstrable belief.[27] The implication of this belief in cosmology is a particular causal structure of the global space-time of the universe; that is, this belief as an act of intentionality cascades down to physical causality.

---

[24] This is a longstanding point made by E. Meyerson in his *Identity and Reality*, London, George Allen & Unwin Ltd., 1964.

[25] See, for example, J. Dancy, *Introduction to Contemporary Epistemology*, Oxford: Basil Blackwell, 1989, p. 116.

[26] There are discussions at present that the universe may not be uniform at large and that the observed uniformity is the result that we are centred in a sort of void.

Another illustration comes from inflationary cosmology: it confesses a belief that there exists a field $\Phi$ (inflaton)[28], which is described through a corresponding theory and which drives the evolution of the universe during the very early inflationary period. This belief coheres (as justification) with another scientific conjecture (belief) that there was a period of evolution of the universe with an exponential growth in time which, in turn, solves some problems of radiation-dominated cosmology[29] and hence makes the so called standard cosmological model even more coherent. One must stress here that all beliefs surrounding the construction of a quite sophisticated theory of the inflationary universe are driven by the hidden desire to explain away the contingent facticity of the initial conditions of the universe as well as its present display. Contingency as eventuality and historicity is not a part and parcel of physics and thus here we observe a certain "pseudo-theological" commitment to overcome the "latent horror of the unique event."[30] A similar situation occurs with the idea of the multiverse. Since no correspondence with empirical reality is possible, all speculations about the multiverse work in the certainty of belief that there is an extended meta-reality which comprises our universe, so that any justification for a theory of such a multiverse can only be based on the grounds of epistemic coherence, which is related to convention at the level of the community of cosmologists. The fact that the idea of the multiverse is driven by a pseudo-theological commitment to justify this universe through the reference to the transcendent can easily be detected by pointing to the by no means rare discussions on how multiverse competes with the idea of creation of the universe *ex nihilo* by God.[31] In the case of the multiverse, in fact, no realistic reference is even required. We deal here with a situation where the mental states (of cosmologists) affect our sense of reality and even contribute to its theory.[32] The idea of the multiverse can be approached from a different point of view if considered phenomenologically as an eidetic variation of the parameters pertaining to the actual universe. This variation takes place within human subjectivity and aims to articulate some apodictic features of that state of affairs which accounts for this actual universe (as a unique event). In this case the invocation of the idea of the multiverse is a legitimate phenomenological procedure in order to reaffirm with a new force the inevitability of the given contingency of this actual universe. But certainly in this case the causation which is implied by the model of multiverse is of a rather mental kind, so that the analysis of conscious states becomes, in a sophisticated way, the datum of scientific facts and cosmology as such becomes a form of phenomenological explication of the working of human subjectivity.

---

[27] Discussing the cosmological principle in close connection with the so called Copernican principle, E. McMullin points out that the Copernican principle has to be understood in terms of what it rejects, namely older teleological beliefs about the uniqueness of the human and the likelihood that humanity has a selected position in space, for example being a cosmic center.( "Indifference Principle and Anthropic Principle in Cosmology", *Studies in History and Philosophy of Science*, v. 24, n. 3, 1993, pp. 359-389 (p. 373)) However the desire to abandon the teleological explanation is itself based in intentionality, rather than any scientifically demonstrable conviction. The indifference postulated by the cosmological principle is indemonstrable because it itself lies in the foundation of the very possibility of scientific demonstration applied to cosmology. Thus it is based in the belief in knowability of the universe which has a different motivation in comparison with that one of teleology (but related to the latter).

[28] In spite of the fact that the hypothesis of this field, its very existence, is very efficient in a qualitative and quantitative modelling of observable phenomena, the physical nature of this field, that is its relation to a certain class of observed particles, remains obscure. This is one of the major points of scepticism with respect to inflationary theories, which has been raised, for example, in the abovementioned paper of Ellis, "Issues in the Philosophy of Cosmology", p. 1210. (See a similar point made in R. Penrose, *The Road to Reality*, London: Vintage Books, 2005, p. 751; also S. Weinberg, *Cosmology*, Oxford University Press, 2008, pp. 202, 217).

[29] These are famous horizon, monopole and flatness problems. See e.g. S. Weinberg, *Cosmology*, Oxford University Press, 2008, pp. 201-208. (See also R. Penrose's *The Road to Reality* in what concerns a certain critique of the inflationary hypothesis, pp. 753-57.)

[30] T. F. Torrance, "Ultimate and Penultimate Beliefs in Science", In ed. J. M. van der Meer, *Facets of Faith & Science*, vol. 1. *Historiography and Modes of Interaction*, Lanham, Maryland: University Press of America, 1996, pp. 151-76 (pp. 166-7).

We see thus that the effectuation of the coherence of epistemic justification in cosmology (which implies a communal or transcendental dimension in cosmology) leads to a different stance on ontological commitment in cosmological discourse. Cosmology is now seen as an enquiry into the condition of appearance of the universe, attaining reality such as it gives itself to be apprehended by human beings and their communities, the very reality of the world in which every sensible entity, astronomical objects, physical bodies including human beings themselves find their place and their meaning. However, this discourse of the appearance does not deal much with a description of what appears at the level of observational astronomy and constructs of theoretical physics, but in a more profound sense with a characterisation of the very conditions (related to the reality of the human) which govern the possibility of appearance (manifestation) of the universe. In other words, it is not, properly speaking, a discourse of the phenomena as such (related to knowledge of facts about the universe), but a discourse of the process of *phenomenalization* of the universe. In a traditional mode of language a discourse pertaining to the *conditions* in which the phenomenon constitutes itself as phenomenon is called transcendental. By becoming more and more conscious of its constraints and possibilities (as related to the place of humanity and its communities in being), the discourse of philosophy of cosmology becomes more and more a transcendental discourse. Correspondingly this discourse reveals itself not only as the discourse of the universe, but as a discourse of human beings.

---

[31] See, for example, discussion of this issue in J. Leslie, *Universes*, London: Routledge, 1989; D. Temple, "The New Design Argument: What Does It Prove?" In *Science, Technology, and Religious Ideas*, ed. M. H. Shale and G. W. Shields, Lanham, Md.: University Press of America, 1994, pp. 127–39; W. R. Stoeger, "Are anthropic arguments, involving multiverses and beyond, legitimate?",in B. J. Carr (ed), *Universe or Multiverse*, pp. 445-57 (455-6); R. Collins, "The multiverse hypothesis: a theistic perspective", in B. J. Carr (ed), *Universe or Multiverse*, pp. 459-480.

[32] This thought was anticipated by Henry Margenau who believed that modern physics could provide an evidence that the nature of its reality is determined not only through causation in empirical reality, but also through intentional acts of thought. In his approach to the nature of physical reality he posed a question: "Is sensed nature the only field of departure or arrival in the process of scientific verification, or will inspection of the eidetic structures of consciousness function in a similar way as dator of scientific fact?" H. Margenau, "Phenomenology and Physics", *Philosophy and Phenomenological Research*, vol. 5, n. 2, 1944, pp. 269-280 (278).

By being engaged in the discourse of the universe as a whole human beings themselves are involved into and subjected to the process of their phenomenalization: on the one hand they take it as their task to control this process through advancing (astronomical) praxis dependent upon their theories; on the other hand the universe remains that overall context and horizon of all horizons which escapes constitution by discursive reason so that it is rather human subjectivity that is constituted by the universe to the extent it cannot comprehend the universe. In this sense cosmology represents not so much that which is *manifest*, that is the universe as such, but *the manifestation*, the manifestation which involves the universe and conscious human beings into the endless constitution.[33] Cosmology reveals itself as a contributor to the phenomenological project, as realization of a transcendental discourse.

**Phenomenological Insight in Cosmology as Explication of the Human**

A phenomenological insight into cosmology makes a reversal of its meaning by shifting the centre of its enquiry from the noematic content (that is related to object) to its noetic pole (related to subject), that is the generating human subjectivity. When scientific reason attempts to enquire into the origin of the universe in an absolute sense the strategy of extrapolation acquires some features of *philosophical* transcendence. But here transcendence is not through physical causation (this would be an impossible break beyond the immanent), but through retaining in the background of all physical representations of the universe, in terms of stages of its evolution, an excess of the universe's intuitive donation in the act of life. Transcendence points towards a simple truth that the reality of the human embodied condition in the universe is not exhausted by those physical aspects which position humanity as temporally and spatially insignificant and hence incommensurable with the universe.[34] Correspondingly cosmology, if it is narrowed to the physical and expressed mathematically, cannot account for the ultimate sense of the universe because it cannot account for the ultimate sense of the human.[35] Since no science can give such an account, the question here is about the boundaries of the human in science. The atomic bomb, for example, being a human creation, characteristically points towards the inhuman, that is to the limits of humanity as such. Thus the atomic bomb as a scientific achievement defines in an apophatic (negative) way the sense of the human. Cosmology plays a similar role: it provides some hints and pointers as to where human comprehension and articulation of the universe becomes paradoxically inhuman (the Big Bang, for example). In this sense the cosmology of the Big Bang becomes a characteristic, although apophatic, explication of the sense of humanity as that formation in being which is looking for its own origin and its own history.[36] A phenomenological insight into the sense of cosmology as explicating humanity's quest for itself thus compensates for the incompleteness of cosmology and reinstates its human creator to its ontological centeredness in disclosing and manifesting the universe.[37] At the same time the limits of physics and scientific philosophy, tested through cosmology, in fact test the limits of humanity to understand its own sense of existence. The incomprehensible universe invokes in the human scientific mind humility and discernment in order to realise the limits of its pretensions to knowledge of the universe which resists disclosure and exceeds the capacity of understanding.[38]

---

[33] C.f. J. Ladrière, *Language and Belief*, Dublin: Gill and Macmillan, 1972, pp. 169, 173, 176.

---

[34] In the context of the so called anthropic inference this was pointed out by M. Bitbol, "From the Anthropic Principle to the Subject Principle", F. Bertola, U. Curi (Eds.) *The Anthropic Principle. Proceedings of the Second Venice Conference on Cosmology and Philosophy*, Cambridge University Press, 1993, pp. 91-100. In a wider philosophical and theological context this excess of humanity beyond nature was discussed in A. Nesteruk, "Theology of Human Co-Creation and Modern Physics", *Mémoire du XXIe Siécle, numéro 3-4. Cahiers transdisciplinaires. Création et transcréation.* Paris: Editions du Rocher, 2001, pp. 163 - 175.

[35] The cosmic environment provides the necessary conditions for human corporeal existence (and this is exactly detected in anthropic arguments) whereas the sufficient conditions do not belong to the sphere of physics and point towards human morality, ethics and some eschatological commitments. See discussion in A. Nesteruk, *Light from the East: Theology, Science and the Eastern Orthodox Tradition.* Minneapolis: Fortress Press, 2003, pp. 200-214.

[36] One can point to similarities between the phenomenology of birth and the aspirations of cosmologists to disclose the sense of birth (origin) of the universe. See A. Nesteruk, *The Universe as Communion*, pp. 247-66. "Is there not, when we read it sufficiently profoundly, an analogy between the deep structure of nature and the structure of human existence as openness, creativity, possibility of accord with the event? The problematic of nature can thus be linked with the problematic of human existence." (J. Ladrière, *Language and Belief*, p. 186.)

Since cosmology assessed, phenomenologically, retrieves the "natural" centring of all non-egocentric tendencies of its world-building narrative in human hypostatic subjectivity, this assessment indirectly calls into question the purported neutrality and objectivity of some of its claims with respect to realities which are beyond empirical verification. It could suggest instead that such "neutral" descriptions of the world operate on the basis of existential concerns formulated in a set of *beliefs* (or myths, which may or may not be related to the *faith* of theology).[39] In this sense the phenomenological stance rejects the view that cosmological knowledge describes the world in itself[40]; rather these descriptions are seen as interpretations that are governed by beliefs which can be qualified as controlled to the extent that they are related to a particular *path* of science in human history.[41] For example, if one is to understand and

---

[37] The idea that a research into the underlying sense of science leads to enlightenment of the ways and *telos* of the human spirit was clearly formulated by many phenomenological philosophers starting from Husserl. Here is a quote from J. Ladrière: "The detail of the life of science must […] be investigated in order to know something of the nature of reason and of its becoming…The destiny of reason is outlined […] in the incessant comings and goings that define the life of science. It is in the patient advance of its history that its finality reveals itself". (J. Ladrière, "Mathematics in a Philosophy of the Sciences", In T. J. Kiesel and J. Kockelmans (eds.), *Phenomenology and the Natural Sciences*. Evanston: Northwestern University Press, 1970, p. 455).

[38] The phenomenological construct of "presence in absence" can be easily applied to cosmology. For example: we see the universe back in time along the so called past light cone, so that the inference about the universe outside this cone can be considered as an attempt to deal with the universe as a whole which is present in its empirical absence. A similar thing can be said if one remembers that according to present-day model the visible matter represents only 4% of the whole material content of the universe. The other 96% (dark energy and dark matter) is postulated in order to balance the model with observations. In other words the universe is present to us through 4% of what is visibly manifested but in empirical absence of the 96%.

[39] A basic and unavoidable structure of any cosmological myth, including its contemporary scientific arrangement is the duality between the factual and empirical on the one hand, and the intelligible (as allegedly stable and underlying) on the other hand. See e.g. J. Ladrière, *Language and Belief*, p. 153.

[40] As it was eloquently expressed by A. Gurwitsch, "the goal of phenomenology is not an exhaustive description of an infinite variety of immanent data, but the investigation of those contexts of consciousness owing to which there is a perceptible world, the universe of physical constructs, etc." (A. Gurwitsch, "Comments on Henry Margenau's 'Phenomenology and Physics'". In L. Hardy, L. Embree (eds), *Phenomenology of Natural Science*, Kluwer, 1992, pp. 35-44 (43-44).)

[41] It is the presence of this concrete path of science which confirms our previous stance on cosmology as the working of constitution, that is a re-enactment of the production of the world. To clarify this point one can quote another paper of J. Ladrière: "The theoretical apparatus is thus not a description in the

explain the past of the universe as constituted through human history, one must conceive it in terms of past possibilities of this history rather than as a defined and finished product. In this case the cosmologist's own historical consciousness is involved in "archaeology" of the universe and, in analogy with historical science, *cosmological discourse reveals itself as a form of consciousness which humanity (as community) has of itself.*[42] By revealing the *telos* in the historical path of cosmological explanation (as related to the representation of the ultimate origin of the universe)[43], phenomenological analysis discloses the hidden "theological commitment" in cosmological research, meaning that the beginning and the end of the universe in human thought is just a mode of this same thought speaking of its own beginning and its own consummation implying a transcendent reference. Seen from a slightly different angle this "theological commitment" corresponds to an attempt to know and see the universe as "all in all", as that primary existential memory in the human constitution which drives cosmological research.[44]

Regardless of possible scepticism with respect to objectivity and neutrality, scientific cosmology remains an extremely important and useful instrument in demonstrating just how human embodied subjectivity affirms itself through the non-egocentric tendencies of its "natural" attitude. The universe that science depicts as something different from us and devoid of our influence and presence, represents in fact the articulated words and thoughts of humanity. By creating a cosmological narrative we affirm ourselves in a non-trivial sense.[45] Indeed, by creating a physico-mathematical narrative cosmologists loose control over the intentions they are driven by, since the introspection upon their creation is not in the focus of scientific enquiry. To understand the "data" lying behind this narrative one must consider it as ordinary sense, as presentation of an entity, supposedly given, and of its properties, it is the characterisation of something which is not a thing, but a *structural path along which a thing comes, from the ultimate horizon of every givenness, to the actual presence in which it is effectively given to apprehension.*" (J. Ladrière, "Physical Reality. A Phenomenological Approach", *Dialectica*, Vol. 43, N 1-2, 1989, pp. 125-39 (138). Emphasis added.)

[42] C.f. R. Aron, *Introduction à la philosophie de l'histoire*, Paris, Gallimard, 1938, p. 80.

[43] See A. Nesteruk, *The Universe as Communion*, pp. 250-254; "From the Unknowability of the Universe to the Teleology of Reason", In J. Bowker (Ed.) *Knowing the Unknowable: Science and Religion on God and the Universe*, I. B. Tauris & Co Ltd, 2009, pp. 63-86 (78-81).

[44] Physical cosmology makes it clear that the world line of the human observer starts at the Big Bang, so that whatever we have on our physical content is directly related to that undifferentiated something lying in the foundation of all possible form of mater in the universe.

[45] See, e.g., B. Swimme, *The Hidden Heart of Cosmos*, New York, Orbis Books, 2005, p. 7; F. Mathews, *The Ecological Self*. London: Routledge, 1991, p. 5; M. Kline, *Mathematics in Western Culture*, Penguin Books, 1977, p. 423.

manifestation of an expressive act, that is to move from their *given meaning* to their *giving meaning*, from their pure phenomenality to the intentional life which generated them. By predicating the evolving universe and attempting to phenomenalise the mystery of its contingent origin, human subjectivity employs that intentionality which effectuates the *telos* of human subjectivity's ever-going incarnation as "coming to presence" assigning thus a dynamic character to personhood's manifestation.[46] As expressed by M. Munitz: "The goals of cosmology are goals of human beings". However, the universe as such benefits from these goals: "Through the measure in which they are reached, the universe becomes understood, perhaps for the first time anywhere throughout its vast stretches in space and time"[47]. By reflecting the goals of humanity, cosmology exhibits the traditional features of all mythologies, namely that the perceptible aspects of the universe are expressed in terms of human social, behaviorial and existential concerns. In this sense the picturing of the universe as a historical process cannot avoid containing erratic facts associated with the human condition, to be more precise with the intimacy of personal communion with the universe and the extent of not being attuned to it. Any imaginable attempt to disregard these facts and assess cosmology only on the basis of law-like ordered concepts would be incomplete and historically inadequate: in this case cosmology would provide us only with a fringe of the universe's phenomenality. The other "part" of the universe's phenomenality which reflects the erratic fact of not being attuned to the universe is rather reflected in poetic and artistic depictions relying on ecstatic act of personal being in the universe as communion. This only confirms an already formulated view that cosmology cannot dispense with anthropology, not only in a high philosophical sense, but in the mundane sense of human affairs.[48] The so called mythological aspect of any cosmology thus naturally arises from the intention to interpret erratic features of the human universe through a reference to the astronomical order and it is this aspect that brings with a new force a "coherence" dimension in its epistemic justifications which was mentioned before: indeed mythologies never present gaps in their "explanations" and are invoked by the communities of adherents on the ground of coherence of their claims.

---

[46] C.f. P. Heelan, "Nature and its Transformations", *Theological Studies*, vol. 33, 1972, pp. 493-502. See also in this context J. Compton, "Natural Science and the Experience of Nature", in J. M. Edie, *Phenomenology in America. Studies in the Philosophy of Experience*, Chicago, Quadrangle Books, 1967, p. 82.

[47] M. Munitz, *Kantian Dialectic and Modern Scientific Cosmology*, *Journal of Philosophy*, vol. 48, N. 10, 1951, pp. 325-338 (338).

[48] As was provocatively conjectured by an author from the camp of the human sciences and arts, we need "a sort of "mytho-scientific, neo-anthropomorphic" theory, one that would stay operational by combining the findings of mainstream science with conjectures based on mythological thought. This type of theory would map features of the universe through images taken from the domain of human social behaviour….Although anthropomorphic theories might not be operational, they can lead to a better understanding of the universe." (Y. Friedman, "Scientific Theory as One of the Fine Arts", *Leonardo*, Vol. 26, No. 4, 1993, pp. 359-63 (361)).

**The dilemma of the object-noematic and act-noetic:**
**the  paradox of subjectivity and cosmology's untestability**

The problem of the interplay between the two dimensions, the natural and the human, in cosmological discourse has it origin in a simple paradox belonging to perennial philosophy but polished within a phenomenological stream of thought, namely, that the universe of the physical cosmology  is produced by means of special intellectual operations rooted in the life world and thus is specifically *different* from the latter, at the same time, the universe apparently proves to be part and parcel of the life world.[49] In other words, on the one hand the physical universe, as an "object" of the natural and scientific constitution, is different from and alien to the human world  thus exceeding the human reality itself; on the other hand, being a mental accomplishment, the universe exhibits itself as cultural reality thus being contained within the life world.  On the one hand, humanity's position in the universe, interpreted through the natural sciences, is such that humanity is  contained by the universe, that is humanity, through physical embodiment, is subjected to the necessities of physics and biology. On the other hand, the universe as being articulated in human perceptual and intellectual experience is contained in human subjectivity so that humanity is treated as a transcendental subject, that is as a free intentional agency, whose personal consciousness cannot be accounted through or reduced to the physical and thus "naturally" belongs to the enquiry of the human sciences.  In spite of this paradox physics sees as its task the development of the theory of the universe  in which conscious humanity would be a product of the universe's law-like-necessities. In this case the paradox would be removed because the foundation of a scientific explanation of the universe (as a mode of humanity's existence) would be part of the universe itself. If cosmology could sustain this ideal then it would prove its right to be qualified as a natural science in the strict sense. In this case the universe will be treated as an external object with no influence from the presence of humanity. Can cosmology, in fact,

---

sustain this ideal? In order to answer this question let us, for a start, analyse one particular aspect in the object-noematic interpretation of cosmology related to its lack of testability.

In the conclusion to his survey of philosophical issues of modern cosmology, George Ellis stated the thesis that uncertainty constitutes a key aspect of cosmology: "scientific exploration can tell us much about the *universe* but not about its *ultimate nature*, or even much about some of its major geometrical and physical characteristics. Some of this uncertainty may be resolved, but much will remain. Cosmological theory should acknowledge this uncertainty."[50] It seems that what is implied here is a distinction between "the universe" as it appears through study and construction by a cosmologist and that which can be termed as the ultimate, underlying sense of its contingent facticity (givenness). The uncertainty of cosmology lies in the fact that it is functioning in the framework of what is already given but, nevertheless, incapable to account for its own possibility as a fact of life.[51] Unlike other sciences (with respect to which one can assert a similar thing, namely that they do not account for their own foundations), cosmology has a particular status because it pretends to deal with the unique and all-encompassing object – the universe as a whole which, by definition, includes not only what is observed here and now (that is in a particular spatial location – home place – and in a particular historical period), but that totality to which human thinking assigns the sense of reality independent of space and time (this totality as a generic mathematical view can have a sense of a Platonic idea, thus being devoid of space-time propensities). Correspondingly, its intended object, that is the universe as a whole, cannot be constituted as accomplished here and now but, on the contrary, represents an asymptotic ideal reached through a self-correcting advance of knowing. In approaching this ideal cosmology invokes many ideas and intuitions (related to constructs and theories) introduced on the level of cosmologists' *intentionality* (supported by beliefs, such as, for example, in explainability of the universe) and not under the pressure of evidence based on empirically accessed physical causality. Thus, naturally, these ideas cannot be tested on the level of physical causality, and are, probably, untestable in principle in a finite span of time.[52]

---

[50] Ellis, "Issues in the Philosophy of Cosmology", p. 1274 (emphasis added).

[51] One can say, by using terms of J. L. Marion, that for cosmology as a scientific discipline it is typical to function in conditions of "positive incertitude", that is asserting things provisionally, which certainty will be complemented by new experimental and theoretical results, or even replaced by a more powerful scientific paradigm. Here lies the paradox of scientific cosmology's functioning: on the one hand it delivers truth about partial objects, but still provides only a provisional truth with respect to the universe as a whole. And it is this positive incertitude which as such forms the condition of cosmology's (and in general science's) progress. In contrast, if one approaches cosmology from a philosophical dimension, it becomes clear that philosophy was "certain" about general aspects of predicating the universe for centuries (remember, for example, Kant's analysis of the cosmological idea in order to realise that its actuality has not been extinguished after nearly a century of scientific cosmology). The perennial problem of the universe's facticity, that is its being, has not been removed from the philosophical scene in spite of desperate attempts of the apologists of the multiverse and the Theory of Everything to remove this problem on the grounds of the immanent science. In this sense philosophy works in the conditions of the "certitude negative" which recognises that there are certain aspects of experience which are subject to discussion and explanation. The universe as donation and event in any life, as the beginning and end, as being and meaning, cannot be explicated in terms of positive definitions. In this sense Ellis' recourse to uncertainty and untestability of cosmology represents an intellectual move towards the "negative certitude". (See J.-L. Marion, "Je suis un sceptique instruit", *Philosophie Magazine*, N. 39, 2010, pp. 58-63, as well as his book *Certitudes négatives*, Paris, Bernard Grasset, 2010).

The accentuation of *uncertainty* and *untestability* by Ellis can give the impression that cosmology, in those parts which do not refer to direct astronomical observations, is not what is usually classed as a "natural science". For the natural sciences it is typical to bring unknown aspects of nature to their explicit presence which is confirmed by observations and tests. Theories typical for the natural sciences aim to refer to empirical reality by means of rules of correspondence, that is through tests and procedures established by scientific community and accessible to the public comprehension on the grounds of common sense. Mathematisation plays an important role in a theoretical advance of the natural sciences, but mathematics, associated by its historical origin with the natural sciences, remains a tool, a method, ultimately created from within the conditions of the life world in order to appropriate this world further through articulation and by means of theoretical thematisation.[53] The aim of the natural sciences is to explain facts by arranging them in the framework of physical causality as a sort of human control.

---

[52] Apart from using the adjective "untestable" 17 times in Ellis' paper, one can find many propositions in which the words "unprovable" or "improbable" are used as characteristics of theories which represent a main stream of cosmology. This wording is used in both cases whether one speaks about the state of affairs in our universe (e.g. "the deduction of spatial homogeneity follows not directly from astronomical data, but because we add to the observations a philosophical principle that is plausible but *untestable*" p. 1226) or if one speculates about the so called multiverse when one attempts to predicate other worlds in terms of that one which is ours (e.g. "multiverse proposals are unprovable by observation or experiment….", p. 1263).

[53] One can remember of Husserl's famous criticism of the radical mathematisation of nature on the grounds that mathematics, as a method providing a certain result of mathematical representation of nature was taken for reality of this nature. Husserl writes: "Mathematics and mathematical science, as a garb of ideas, or the garb of symbols of the symbolic mathematical theories, encompasses everything which, for scientists and the educated generally, *represents* the life-world, *dresses* it up as "objectively actual and true" nature. It is through the garb of ideas that we take for *true being* what is actually a *method* – a method which is designed for the purpose of progressively improving, in infinitum, through "scientific" predictions, those rough predictions which are the only ones originally possible within the sphere of what is actually experienced and experienceable in the life-world." (E. Husserl, *The Crisis of European Sciences and Transcendental Phenomenology*, Evanston, Northwestern University Press.p. 51-52.)

If now cosmology is perceived (in the spirit of Ellis) as an enterprise which bases itself in non-testable assumptions, there is a question that arises on the general sense of this enterprise and validity of its epistemic claims. Do cosmological theories indeed deal with physical reality, posed as existing in itself, so that cosmological constructs provide a set of convergent approximations to it? Or does, alternatively, cosmology, being a mode of cultural activity, just create an intellectual environment with a corresponding narrative on humanity's place in the universe without any ontological commitment? It seems reasonable to conjecture in this case that any narrative about the universe is based on empirical extrapolations and intellectual conjectures which in some cases are untestable *now* and some are untestable *in principle*. Nevertheless, these untestable conjectures contribute to the wholeness of the narrative thus becoming indispensable constituents of the idea of the universe. The natural origin of these conjectures is related to their historical contingency which entails cosmology's *positive incertitude* and thus the "natural" presence in it of some untestable or eventually falsifiable and theoretically redundant statements.[54]

The problem is that by its definition *the universe as a whole* cannot be thought as a contingent "formation"(unless in a theological sense as created out of nothing) because it assumes a spatio-temporal totality which as such exceeds any contingency in terms of space and time. This invokes a conflict with the historical contingency of cosmology as activity of knowing. The question is: can the historically contingent discipline of cosmology (as related to contingent events of human subjectivity) pretend to encompass the universe as a whole which, by definition, exceeds all spatially distinct realms and eras and thus all aspects of contingent givenness.

---

[54] One can generalise this thought by asserting the historical contingency of the conditions of disclosure in cosmology, that is of the historical relativising a-priori as relative to a certain situation of cosmology changing from one step to another of its history. It is this positive incertitude of cosmological research that guarantees its progress and at the same time makes it radically different in comparison with a "negative certitude" of philosophical insights in the topic of the universe or the world. According to the spirit of the volume on transcendentalism in modern physics (M., Bitbol, et al., *Constituting Objectivity*, p. 13) the historical relativising of the conditions of cosmological knowledge could extend the Kantian understanding of constraints on knowledge, as transferred from a-historical a-priories to ever corrigible conditions of constitution of reality related to the evolving human condition.

The historical contingency of cosmological facts can be easily illustrated. By studying astronomical objects through their image in the two-dimensional celestial sphere we study de facto *free phenomena*, that is those which contain in themselves or establish from themselves the reasons for their being given to us. Their contingent facticity consists in that they are related to historically concrete and individual acts of consciousness. Their appearance is free from any underlying causes in the following sense. Astronomical observations can be interpreted as caused by the contingent factors of human history; however, one cannot construct a trans-historical "trajectory of knowledge" as if it would be driven by a sort of law in the "space" of all possible knowledge and which as a potential possibility led to the given outcome of observations. Or, in different words, one cannot assume the existence of a universal "observer" which could anticipate the path of knowledge independently and apart from the acts of knowing as their first time happening. Knowledge and experience are accumulated in consequent observations in time from different perspectives where the conditions of knowledge, the very fashion in which this knowledge is collected, are not controlled by us; we do not know all factors which influence the course of that existential manifestation which is called "knowledge".[55] This rephrases the sense of the contingency of knowledge as that process which is not exhausted or driven only by its subject matter: the choice of what to observe (or whether to observe or not) and of the notions for the description of the universe is based not in fully understood physical laws and interconnections, but remains hanging within the *free intentionality* of researchers. Correspondingly the untestable conjectures naturally contribute to this intentionality. Untestability of some conjectures about the universe thus reflects the contingency of the conditions of constitution of cosmology's "object", the process which is not one-to-one related to the causal developments in the object of knowledge and sources of experience.[56] The uncertainty of cosmology witnesses to the fact that the *intentional* acts of cosmological research which allegedly attempt to reveal causality in physical processes do not lead to the fulfilment of intentions and this constitutes an irreplaceable and unavoidable feature of any attempt of knowing the universe as a whole.

---

[55] Up to some extent this is true even in the case when the experiments are anticipated through theory and thus planned in advance within a given paradigmatic constraint. The resistance of the universe to its disclosure makes these experiments amendable and corrigible because their outcomes are still contingent.
[56] One implies here a change in the conditions of constitution which are not related to the instrumental requirements and means of accessibility. As an example one can refer to changes of theoretical paradigms which have philosophical motivations; a typical example follows from a philosophical desire to explain away the problem of the contingency of the initial conditions which results in inflationary cosmological scenarios or ideas of the multiverse.

There are some elements in the sociology of cosmology which can be treated similarly to the work of a proper historian.[57] The cosmological narrative can receive a historical interpretation which vindicates in a different way the presence of untestable elements in cosmology. For example, by studying the history of astronomy (as a certain introduction into cosmology proper) one must take a neutral position with respect to epistemic claims of Ptolemy or Giordano Bruno. Certainly from a modern view their cosmologies were incomplete and contained untestable elements. But a *historical perspective* stops us from such an evaluation, accepting, rejecting or amending their epistemic claims concerning the knowledge of the universe unless we evaluate their theories through the eyes of the present era, and our understanding of history is fundamentally conditioned by present scientific views. The same perspective can be applied to modern cosmology. Indeed, if cosmology is seen as an ongoing narration (as intrinsically historical and hence contingent) about the universe, then the presence in its structures of untestable elements as well as overall uncertainty cannot be used to qualify for the epistemic unfoundedness of cosmology. On the contrary these untestable elements should be seen as *cultural artefacts* which must be accepted as existential events, that is as historically motivated actions. Then one must abstain from judgements on the truth of cosmology's epistemic claims and keep neutrality with respect to them. In this case the presence of untestable hypotheses in cosmology can be treated as inevitable. From the point of view of physics this can be a misfortune, but seen philosophically this turns out to be a characteristic of the human condition in which knowledge of the universe takes place.

The narrative about the universe which contains untestable propositions, being historical, is subject to ongoing change and renewal (in terms of the research practice this means that theories change rapidly and the survival of theories against the flow of data (that is testable propositions), let's say within two years, is treated as its success). In cosmology, especially related to ultimate questions (such as the origin and end of the universe, as well as multiverse), this renewal does not only follow the logic of revealing new causal connections among facts of nature (because nature is simply inaccessible), but is driven by intentionality which reflects the aspirations and advances of the human spirit, its "infinite tasks" in grasping the sense of existence.[58] In this sense some conjectures about the untestable aspects of the universe as a whole (being acts of free human will rather than necessities imposed by the logic of nature) represent *existential events* (manifestations and assertions of that which does not present itself through physical causation, that is elements of intentions, motivations and goals of a historical cosmologist-actor), through which personal experiences and intuitions of belonging to the universe attempt to express themselves outwardly in the scientific narrative.[59] Cosmological narrative thus appeals to the idea of a wider cosmic context, beyond this planet, by shifting home places not only in the visible space, but beyond it, towards the intelligible universe. This happens at the expense of leaving out the sphere of the empirical and sensible experience and making a transition to stable and allegedly non-corruptible intelligible entities which are inevitably beyond the reach of any straightforward testability. In some extreme versions of such a narrative its apologists appeal for the removal of all "human baggage" behind the underlying theories.[60] It then seems obvious that in this tendency the grain of untestability and the lack of correspondence with physical reality is posited in the very inception of the non-egocentric aspirations of cosmology. Put another way, untestability becomes an explicit manifestation of the incommensurability between man and the universe resulting in an unavoidable positive incertitude of cosmological knowledge. It is this incommensurability that is characteristically revealed through an attempt to construct a full computational synthesis of the universe.[61] The totality of the universe and its actual infinity, being the source of the incommensurability, is, however, not beyond its reach in the sense of manifestation. To the extent, therefore, that a particular cosmological theory is functioning adequately even in the conditions of untestability one can regard the structure which it articulates as constituting the intelligible pattern of the universe as a whole. Remembering, however, that this intelligible pattern never exhausts the ultimate sense of that which is intended as the universe, the untestability demonstrates itself as the constituting element of the apophatic intelligibility of the infinite universe.

---

[57] C.f. Ellis, "Issues in the Philosophy of Cosmology", p.1221.

---

[58] Husserl defines humanity as mankind with infinite tasks as a carrier of philosophy, conceptions of ideas through which "man gradually becomes a new man", man who "lives toward poles of infinity." Infinite tasks are associated with the theoretical attitude to the world. In this sense science itself signifies the idea of the infinity of tasks. (See E. Husserl, *The Crisis of European Sciences and Transcendental Phenomenology*, pp. 277-279.)

[59] The creation of a cosmic narrative implies a deviation from the egocentric intentionalities which could be criticised by the philosophers of existence, who believed that it was wrong to interrogate the sense of the given facticity; on the contrary their philosophy takes this facticity for granted as an undeniable premise of all other enquiries.

[60] See in this respect M. Tegmark's paper "The Mathematical Universe", *Foundations of Physics*, vol. 3 8, n 2, 2008, pp. 101-150, in which the author attempts to advocate an extreme view of mathematics as an underlying structure of reality, that is as reality itself, which is stripped off of all aspects of the human presence. However the hypothesis of the mathematical universe suffers from not being placed in the context of a serious philosophical discussion on realism in mathematics (such as given in, for example, M. Balaguer, "Realism and Anti-Realism in Mathematics", in A. D. Irvine (ed.) *Philosophy of Mathematics*. Amsterdam: Elsevier, 2009, pp. 35-101), not to speak of a phenomenological stance on mathematics. In addition one must say that the claim that the universe is mathematical, which in the paper of Tegmark is tantamount to assertion of the universe's epistemic exhaustability (that is knowability), contrasts with understanding that mathematical concepts have no intuitive content and are very poor in donation thus leaving behind all aspects of experience of the universe which, on the contrary, are so powerful in donation, that they block any pregiven structure of the discursive reason and constitute it to the extent this reason cannot cope with the saturation of the intuition. (See in this respect J. L. Marion *Being Given*. Stanford University Press, 2002, pp. 179-247. In simple words, all extreme views of mathematical realism assume that if the universe is only physical, then it must be mathematical. However, it is here, that philosophers could raise a serious doubt on whether the physical and hence mathematical representation of the universe exhausts the sense of its reality as perceived by human beings.

Cosmology cannot avoid dealing with hardly testable conjectures just because it is not an experimental science: according to Ellis' theses "The universe itself cannot be subjected to physical experimentation" and "The universe cannot be observationally compared with other universes".[62] In this sense the presence of speculative mathematical elements which cannot be directly related to empirical reality must be accepted with a sort of humility, and in no way as the end of realistic commitment in research, assuming that reality stands here not for a pre-existent antecedent entity, but as being unceasingly constituted. The persistent absence of the physical correlate to that which is intuited through mathematics rather represents an invitation to continue the scientific quest when the reality of what is called the universe is withdrawn from a simplistic empirical grasp. The presence of unattained goals, as imaginatively projected standards, encourages and activates in a scientist a different intentionality which is based on existential inspirations and not fully articulated insights.[63] The initiation of this intentionality allows a philosopher to appropriate cosmology not as that type of knowing which delivers ultimate truth about reality but rather as a particular *account* of the human encounter with the universe. In this case the demand for cosmology follows from existential orientation, that is from asserting  the sense of human existence and its *telos*.[64]  Correspondingly the introduction of untestable conjectures points rather to mechanisms of functioning of human subjectivity when it faces phenomena which exceed their capacity of constitution. This in turn contributes to a generic thesis that the study of the universe contributes toward the study of man.[65] Cosmology can be seen as revealing the structures of human subjectivity in the case of incommensurable phenomena  thus explicating the general functioning of this subjectivity as embodied in the universe.[66]

---

[61] The fact that cosmology involves a "computational synthesis" of the observable astronomical phenomena implies that physical objectivity of these phenomena as well as of the universe as a whole cannot be tantamount to an ontology of some independent reality. The possibility of a complete mathematical reconstruction of such an ontological reality would ascribe to the human mind excessive intellectual capacities which transcend its finitude following from the limits of embodiment. Correspondingly, according to H. Weyl's qualification, cosmology here concedes to idealism (we would say transcendental idealism) in the sense that "its objective reality is not given but to be constructed…, and that it cannot be constructed absolutely, but only in relation to an arbitrarily assumed coordinate system and in mere symbols". H. Weyl, *Philosophy of Mathematics and Natural Science*, Princeton University Press, 2009, p. 117.

[62] Here is the full version of his theses A1 and A2: (A1) "The universe itself cannot be subjected to physical experimentation. We cannot re-run the universe with the same or altered conditions to see what would happen if they were different, so we cannot carry out scientific experiments on the universe itself"; (A2) "The universe cannot be observationally compared with other universes.
We cannot compare the universe with any similar object, nor can we test our hypotheses about it by observations determining statistical properties of a known class of physically existing universes."
G. Ellis, *Issues in the Philosophy of Cosmology*, p. 1216.

---

[63] These existential inspirations initiate generic cosmological mythology which has already been described in terms of cosmogenesis, that is how the world was made, how the present universe which stands before our eyes developed from what went before, from the non-universe, the formless. (Cf. J. Ladrière, *Language and Belief*, p. 153.)

[64] This  thought was expressed in numerous ways by philosophers and scientists. See in this respect, for example,  E. Minkowski, "Prose and Poetry (Astronomy and Cosmology)". In T.  J. Kiesel and J. Kockelmans (eds.), *Phenomenology and the Natural Sciences*, pp. 239-47 (244); M. K. Munitz, "Kantian Dialectic and Modern Scientific Cosmology",  p. 338; P. Brockelman, *Cosmology and Creation. The Spiritual Significance of Contemporary Cosmology*, Oxford University Press, 1999, p. 42; J. Primack, N. Abrams, *The View from the Centre of the Universe*, London: Fourth Estate, 2006, pp. 280-290.

[65] C.f. : "By studying  the forms of objectivity assumed to be present in nature, one can, however, infer the forms of subjectivity that are presupposed. Inquiry of this kind must proceed according to phenomenological method, the purpose of which is to uncover the noetic pole constitutive with the noematic pole, of the noetic-noematic (subject-object) intentionality structure within which the form of the question and the form of the answer mulually determine one another." (P. Heelan, "Nature and its Transformations", p. 486);  or "We can enhance the sense of ourselves, as we've been successfully doing with our other senses, by means of a scientific but nevertheless metaphorical telescope- a new cosmological lens through which we can see how the expanding universe really works and how astoundingly special our place is in it" (J. Primack, N. Abrams, *The View from the Centre of the Universe*,  p. 282). Placed in a cosmological context the following quotation from F. O'Murchadha points to the same thought: "The otherness of nature is beyond us, beyond our humanity, beyond history. Yet it is in a  place carved out of nature that we dwell. In dwelling – as ethical being -  we become  what we are. Hence, *how our dwelling is conceived, how the relation of the history of this dwelling to the natural environment, through which it is carved out, occurs, relates directly to who we are*." O'Murchadha, F.,  "Nature as Other: Hermeneutical Approach to Science", In B. E. Babich et al  (eds.) *Continental and Postmodern Perspectives in the Philosophy of  Science*, p. 189).

[66] One can make a parallel between our phenomenological interest and a similar approach to philosophical issues of physics which is developed in the movement called "formalised epistemology". (See, for example, F. Bailly, "Remarks about the Program for a Formalized Epistemology", in M. M. Mugur-Schächter, A. van der Merwe (eds.) *Quantum Mechanics, Mathematics, Cognition and Action. Proposals for a Formalized Epistemology.* Kluwer Academic Publishers, 2002, pp. 3-8). However, the reduction performed in the "formalised epistemology" does not reach the goal of this research which aims not only  to explicate the cognitive structures underlying science, but also attempts to relate them to fundamental existential conditions predetermining their facticity.

By making explicit the workings of human corporeal subjectivity cosmology places itself within the cultural world thus exhibiting some features of the cultural and human sciences, in particular, along the lines argued by Husserl: namely that the cultural or human sciences reveal themselves as all-encompassing, since they also comprise the natural sciences and mathematised nature since it is itself a mental accomplishment, that is, a cultural phenomenon.[67] Cosmology acquires the meaning of a cultural science in the sense that it deals not only with the disclosure of "objective" reality of the universe (as humanity's natural environment) but encodes human aspirations to disclose the sense of its place in the universe, for example, by measuring the universe through the standard of human life.[68] It is in this sense that cosmology exhibits itself not as a monological questioning of the universe as if it was out there, but rather as a dialogue with the universe as a noematic pole through which humanity enquires of itself.[69] The fact that this noematic pole is not fixed and escapes its ultimate grasp constitutes a particular feature of the cosmological enquiry in which persons as centres of disclosure are formed, as disclosers, to the extent the universe discloses itself. The ego's subjectivity is evolving through the invitation by the universe to disclose it in certain limits. Thus there is no independent object of disclosure, independent of those who are participating in the "dynamics" of this disclosure.[70]

---

[67] E. Husserl, *The Crisis of European Sciences*, p. 237. However the converse is not true, that is the cultural sciences cannot be given a place among the natural sciences. See in this respect A. Gurwitsch, *Phenomenology and the Theory of Science*, pp. 148-9.

[68] Here it is appropriate to point to a certain similarity between the phenomenology of birth and theories of the beginning of the universe as related to a general problem of consciousness' facticity. (See A. Nesteruk, *The Universe as Communion*, pp. 247-250.) In the natural attitude the same problem can be addressed under the name of genetic similarity between the *biology* of birth and the stages of development of the universe (See A. N. Pavlenko, "A Place of 'Chaos' in the New World 'Order'", *Voprosy Filosofii*, N9, 2003, pp. 39-53 (47-48) (in Russian); "Universalism and Cosmic Harmony: A Principle of Genetic Similarity", *Skepsis*, XV(i), 2004, pp. 389-401). The views on the universe were always important for communities to draw some remote expectations having ethical character: ethics depends on the idea of environment and its developmental perspective. (See in this respect F. Mathews, *The Ecological Self*. London: Routledge, 1991, pp. 3-6.)

[69] The constitution of one's ego through knowledge so that the "object" of this knowledge is formed together with its subject represents a crucial feature of the human sciences. According to M. Bakhtin, the object which is studied in the human sciences belongs to the same realm as the subject who studies, and thus it is no less active than the knowing subject (M. Bakhtin, *Aesthetics of Verbal Creativity*, Moscow, Iskusstvo, 1979, p. 349 (In Russian)). In this context it is interesting to make a reference to J. A. Wheeler who affirmed a similar thing with no recourse to the human sciences. For example: "In giving meaning to the universe, the observer gives meaning to himself, as part of that universe" (C. M. Patton, J. A. Wheeler, "Is physics legislated by cosmogony?", in R. Duncan, M. Weston-Smith (Eds.) *Encyclopaedia of Ignorance*, Oxford: Pergamon, 1977, pp. 19-35 (31)). That is, by disclosing the universe the observer forms its own structures of subjectivity which apprehend the universe.

[70] The opposite would imply sheer idealism, resulting in the universe to be an intelligible entity graspable through the already formed intellect. The universe would be thought as pre-existent, but intelligible,

Correspondingly one intuits the universe as related to the continuity of human conscious experience of embodiment in the universe. Seen in this way, cosmology could acquire a teleological sense if one relates it to humanity's "infinite tasks" of dealing with questions of the beginning and the end of the universe as having connotations with enquiries into the sense of its own beginning and its own consummation. It is because of the infinitude of these tasks that human subjectivity while attempting to comprehend the sense of the beginning and the end of the universe, is forming and comprehending itself. It is through the unlimited donation of the universe (as the field of possibilities) that humanity awakens to its own finitude in spite of a potentially infinite consciousness (based in the infinite human will), that is to the issue of the "beginning" and "end" of consciousness itself, that is, its contingent facticity. This sense of finitude entails another mode of incommensurability, that is the infinite can only be represented through symbols (apophaticism) which by their origin function not on the level of physical causality.[71] In this sense cosmological theorising as an ongoing symbolising the universe contains in itself logic and necessities which are not directly related to its subject matter, but rather to the sense of goals of humanity itself.

**Cosmology and human will**

In the background of what we have said so far there still remains a question (related to Ellis' comments on the presence of untestable and unprovable assumptions in cosmology) as to why in spite of the a-priori understanding that cosmology will never achieve the fullness of explanatory power and adequacy with truth, the urge for cosmological search and narration continues. Indeed, even for a reader unexperienced in all the subtleties of the methodology of science the claim that cosmology is based in

---

Platonic-like entity. Its grasp, would imply a sheer mysticism as the communion with the realm which is beyond the empirical (this mysticism is similar to that one which is envisaged in the Platonic philosophy of mathematics). Even if one would allow for a complete computational synthesis of the universe, this would, as we mentioned above, exceed human finite capacities limited to the conditions of embodiment. The desire to know the universe in its totality (as "all in all") brings to mind eschatological connotations as if the thinking of totality of the physical universe is equivalent to the anticipation of the eschaton in which the overall transfiguration of the world and human beings will allow one to see the universe from the perspective of the trans-worldly existence.

[71] H. Weyl linked the human longing for the sense of the infinite with that moral and human standard by which all human deeds are judged: "…mind is freedom within the limitations of existence; it is open towards the infinite.… The completed infinite we can only represent in symbols. From this relationship every creative act of man receives its deep consecration and dignity." (H. Weyl, "The Open World: Three Lectures on the Metaphysical Implications of Science". In his *Mind and Nature*, Princeton University Press, 2009, pp. 35-82 (82)).

a fundamental uncertainty must make a staggering impression, that cosmology formed an exotic set of trans-scientific ideas and intuitions which, by virtue of popular science and mass culture, acquired the status of a stable social *belief*. However, cosmology is not a new mythology[72] or a kind of cosmic philosophy[73] (based on wishes and remote expectations of community), it is not a sheer imagination, but has its own logic and drive, which, reflects the sense and value (as well as *telos*) of communion with the universe conditioned by necessities of nature and, at the same time, pertaining to human freedom. This communion is rather a state of *apatheia* as transcending over the natural causes which are beyond human control and. At the same time, communion with the universe is the immanence to the universe through the sheer fact of human presence, through which the universe is transformed by human will, its *energeia* and operations of cognition. Thus the persistence of cosmological research comes not only from the logic of cosmological research, but from other factors originating in the human condition.[74] Indeed, finite human beings, because of their paradoxical standing in the universe are not content with the presence of things in the universe as they are given in their empirically contingent facticity. Cosmologists, by invoking the idea of the universe as a whole, manifest their desire to understand the meaning of finite things (let us say astronomically observed objects or earthly phenomena) not only through their *nature* (that is through that which is subjected to physical causation), but through the *purposes and ends* of these things as they stand with respect to the universe as their ultimate foundation. But this intentionality directed into the foundations of the very facticity of things is not what can only be expressed outwardly in terms of physical causation (and thus subject to tests); it is sustained by humanity's aspirations not only to be commensurable to the universe but, in fact to be above it, to transcend it and thus to encompass it through the power of intellect rooted in human will. It is because of its paradoxical position in being which causes existential discomfort[75] that humanity appeals to the idea of the universe as a whole as an alternative to being contained by finite natures, that is being comprehended only as an "object" among other objects in the universe. Existentially it does not want to be manipulated through circumscribability and individualisation which are inherent in spatio-temporal forms of the finite cosmos and correspondingly long for the truth of their existence *in* the space-time rubrics of this universe as if it is not *of* the universe as it appears to us. Here humanity wants to recognise things not according to their compelling givenness, but as results of humanity's *free will* realised in its intentionality. This naturally leads to the transcendence of the empirical and the invocation of intelligible entities (sometimes untestable and unprovable) which serves as a pointer and invitation to further research rather than its end and ultimate certainty.[76] The presence of agents with free will in the universe imposes certain constraints on the nature of the universe: it must contain the necessary conditions for them to exist[77] (or, as argued elsewhere, the universe must be moral[78]).

---

[72] Here one must agree with Ellis that the fact that cosmologists write about contentious issues in cosmology "is proof that they consider it meaningful to argue about such issues" because their quality emerges naturally from knowledge of the physical universe (Ellis, "Issues in the Philosophy of Cosmology", p. 1272).

[73] Here one means the so called cosmic philosophies and ideologies of the ancient past which aspired for humanity to be dissolved in cosmic immensities and which were strongly dismissed by the ecclesial authorities as pagan and gnostic.

[74] J. Moltmann, by formulating the quest for the sense of cosmic utopia which creates in us the interest to know the universe, summarises that all that which the vast science-fiction tells can be reduced to the following: the infinite survival of humankind and unlimited development of human consciousness (J. Moltmann, *Science and Wisdom*, Minneapolis: Fortress Press, 2003, p. 72.)

---

[75] E. Fromm, for example, speaks about existential and historical dichotomies in man. Existential dichotomies are related to man self-questioning the sense of its own existence: " Man is the only animal which can be bored, that can be discontented, that can feel evicted from paradise.  Man is the only animal for whom his own existence is a problem which he has to solve and from which he cannot escape." These are dichotomies about life and death,  loneliness and relatedness, individuality and sociality. Because of their free will man can attempt to annul historical contradictions through their actions, but it is futile to overcome existential contradictions, man remains dissatisfied, anxious and restless. Then "there is only one solution to his problem: to face truth, to acknowledge his fundamental aloneness and solitude  in the universe indifferent to his fate, to recognise that there is no power transcending him which can solve his problem for him" (E. Fromm, *Man for Himself. An Enquiry into the Psychology of Ethics*, London, Routledge and Kegan Paul, Ltd., 1967., pp. 41-44).

[76] As eloquently expressed by H. Küng there is no intellectual compulsion in  questions beyond empirical reality, but *freedom* dominates in them (H. Küng, *The Beginning of all Things*, W. B. Eerdmans Publ. Comp., 2007, p. 78). Here comes to mind an analogy with the Kantian aesthetical ideas which, being qualified as inexponible presentations of imagination function according to *"free play"* (I. Kant, *Critique of the Power of Judgement*, § 57, Comment 1.  Tr.  P. Guyer, E. Matthews, Cambridge University Pres, 2000, p. 219). The analogy is that the question about the invisible foundation of the universe falls under the rubric of aesthetical idea, rather than rational idea.

[77] This argument corresponds to what is generally called  anthropic inference in cosmology, namely a very delicate interplay between the physical and biological parameters of human existence and large-scale properties of the universe as well as fundamental physical constants. It is important to realise that this inference does not account for the facticity of humanity's existence, because it does not cover the realm of sufficient conditions which  belong to the sphere of human morality and conscious will. Indeed the technological advance of humanity threatens its local survival on the planet without affecting the global physical properties in the world. This implies that the actual presence of humanity in the universe as an ongoing event is determined by human wisdom and morality rather than simply by the cosmic conditions. (See, e.g., J. Leslie's *The End of the World: The Science and Ethics of Human Extinction.* New York: Routledge, 1996; Rees, M. *Our Final Century, A Scientist's Warning: How Terror, Error, and Environmental Disaster Threaten Humankind's Future in This Century – On Earth and Beyond*. London: W. Heinemann, 2003.)

[78] The assertion of  the morality of the universe advocated in the book by N. Murphy and G. Ellis *On the Moral Nature of the Universe*, Fortress Press, 1996 (see in particular p. 207) does not have any

The perception of cosmology as that block of insights which involves deeply human anxieties and correspondingly persists as an existential quest, invokes a different stance on the ontological commitment exercised by advocates of cosmology, all those who are engaged in its popularisation and adoration and who usually claim, that whatever is theoretically and mathematically formulated, is physically real and true, although non-observable and untestable.[79] The countersense which is put forward by the human sense of cosmology doubts not the legitimacy of the cosmological narrative (comprising theories of non-observable entities) *per se* but the validity of epistemic justification adopted for its realistic interpretation.

Indeed, if many cosmological hypotheses and inferences are not testable, that is the correspondence principle between theory and empirical reality as epistemic justification does not work, there is a way of interpreting cosmological propositions about the non-observable and invisible by assigning to the universe the sense of a mental accomplishment but achieved through the idea of *coherence*[80], where "coherence" stands most of all not for the clarity of theoretical explication and cohesion of mathematical calculations, but for the "collaborative agreement"[81] among cosmologists. It is these cosmologists who, by exercising their will, effectively *hypostasise the notion of truth* related to the universe and postulate the ways of epistemic justification which lead to it.[82] In this case the implied truth of cosmology

---

straightforward scientific reference, for "free will" as well as the very facticity of consciousness in the universe cannot be accounted through any reduction to the physical or biological. One can add to this that the very existence of cosmology as a free and creative questioning of the universe is thus inherent in the fact of free human choice to explore the world at large. The initiation of cosmology lies in the freely made decision to act and exceed the limitedness of the empirically given. And this free decision as such is not subject to a scientific account.

[79] As an example one can point to M. Tegmark who conjectured a principle of "mathematical democracy" according to which whatever is mathematical is also physical. (M. Tegmark, M., "Parallel universes", pp. 480-485). The reader should remember, however, that the main issue here is whether the mathematical exhausts the whole of reality. One can adopt a different view on mathematical models of the universe regarding them as related not to one and the same physical original. In this case a cosmological model consistent from the mathematical point of view can give images of that which cannot be physical in this universe. In spite of this all mathematical models being created in this universe contribute in a sort of way to its articulated content.

[80] N. Rescher, *The Coherence Theory of Truth*, University Press of America, 1989, pp. 318-9.

[81] Ibid, p. 333

[82] See a book of J. Bowker *The Sacred Neuron*, London, I. B. Tauris, pp. 118-148, in which the author persuasively argues on the importance of coherence considerations in science (and religion) as a different form of justification in comparison with the correspondence principle. On the limited application of the idea of coherence of epistemic justification in cosmology see A. Nesteruk, *The Universe as Communion*, pp. 244-46; "From the Unknowability of the Universe to the Teleology of Reason", pp. 71-75;

cannot be an ontological truth (that is physical truth as allegedly existing in itself) but is a human-dependent constitution of truth possessing the qualities related to the corporeality of human beings.[83] In this interpretation many cosmological constructions naturally acquire the status of coherent mental accomplishments (based in beliefs) whose truth (being historically contingent) contributes towards the spiritual goals (*telos*) of community but obviously does not exhaust them. However, the question of whether the locally established truths are subject to convergence to ultimate truth remains beyond scientific scope and represents in turn a belief motivated by trans-scientific convictions. In this option the validity of cosmology's claims is dictated not by a direct reference to reality but through the adoption of a consistent and creative set of beliefs which themselves constitute the sense of reality, although contingent as related to the goals of community of cosmologists.[84]

Since in thus treated cosmology the universe appears to be a collaborative construction, its knowledge cannot be treated as independent of human insight, so that cosmology's alleged status of following the standards of a natural science (as that in which the "object" of study can be entirely separated or detached from the subject) is not achievable. Cosmology, in contradistinction with astronomy and astrophysics[85], is rather the "universology"[86] which deals with a single, unique totality of all, which not only cannot be treated as an object and hence subjected to experimentation, but also cannot be made devoid of the delimiters of human insight. This, as we mentioned before, implies that human beings, as part of the universe, cannot position the universe

---

"Transcendence-in-Immanence: a new phenomenological turn in the dialogue between theology and cosmology", In V. Porus (Ed.) *Scientific and Theological Epistemological Paradigms*, Moscow, St. Andrew's Biblical and Theological Institute, 2009, pp. 35-62 (48-61) (in Russian); as well as in the English version of the same paper "Transcendence-in-Immanence in Theology and Cosmology: a New Phenomenological Turn in the Debate", *Studies in Science and Theology*, vol. 12, 2010, pp. 179-198.
[83] Cf. M. Bitbol, et al. (Eds.). *Constituting Objectivity*, p. 4.
[84] This allows one to make a certain analogy between the forming of sense in cosmology and theology: indeed theology forms its sense of truth not through empirical references to the Divine, but through the experience of God as elaborated and established ecclesial agreement. ( See, for example, A. Nesteruk, *The Universe as Communion*, pp. 244-46.)
[85] G. Ellis underlines the essential characteristic of cosmology's proper subject matter: "..if we convince ourselves that some large scale physical phenomenon essentially occurs only once in the entire universe, then it should be regarded as part of cosmology proper; whereas if we are convinced it occurs in many places or times, even if we cannot observationally access them…then study of that class of object or events can be distinguished from cosmology proper precisely because there is a class of them to study", ("Issues in the Philosophy of Cosmology", p. 1219). This careful distinction is related to the universe as a whole and makes a clear-cut demarcation line between cosmology *proper* and other celestial sciences like astronomy and astrophysics.
[86] S. Jaki, *Is There a Universe?* Liverpool University Press, 1993, pp. 1-2.

as a whole in front of their consciousness, unless as a mental abstraction.[87] If such a mentally constructed universe nevertheless were to be identified with the physical totality, this would imply a sort of impossible transcendence of the actual physical universe as if one were able to "look at it" from the outside and hence transcending one's embodied existence.[88] The inseparability of humanity and the universe as their consubstantiality entails that all speculations about other worlds remain intrinsically immanent, being noematic correlates of embodied subjectivity which is an irreducible element of being of this universe.[89] Thus the universe as an intentional correlate of cosmological consciousness represents a mental accomplishment and cultural achievement[90] exhibiting features of the sciences of human affaires.[91]

**The explication of the interplay between natural and human sciences in cosmology**

Let us articulate further the sense of the interplay between the elements of human and natural sciences in cosmology. Cosmology is a scientific activity of human beings: it is because of this that in its constitution it is a human science in a trivial sense.[92] This claim comes from a noetic pole and implies that the epistemic and socially significant achievements of cosmology just are cognitive and manipulative achievements of human beings.[93] To say that "cosmology is human science" is to say that the doing of cosmology is an existential characteristic of human beings, their mode of being-in-the-world.

---

[87] This would correspond to a Platonic treatment of the construct of the universe as an idea. In this case cosmology were to face, in analogy with the general Platonic stance in the philosophy of mathematics, a serious problem of justifying the interaction between the universe as an intelligible entity and its empirical appearance to embodied consciousness, the interaction which would imply a sort of mystical communion. (See in this context a nice discussion on the status of mathematical objects and their knowability in R. Tieszen, *Phenomenology, Logic, and the Philosophy of Mathematics*, Cambridge University Press, 2005, pp. 46-68.)

[88] C.f. G. Marcel, *Being and Having*, London: Collins, 1965, p. 24.

[89] One can argue that the very process of invocation of other worlds is reminiscent of that which phenomenology calls "eidetic variation". One subjects the physical parameters of the universe to a sort of variation whose ultimate goal is to establish the stability or the *eidos* of the actual universe.

[90] C.f. A. Gurwitsch, *Phenomenology and the Theory of Science*, pp. 44-45. See also E. Husserl, *The Crisis of European Sciences and Transcendental Phenomenology*, p. 227 on the "nature" as correlate of a universal abstraction.

[91] A similar observation, with no recourse to phenomenology and the concept of intentionality, has been made by a Russian philosopher V. Rosin. See, for example, V. M. Rosin, *Types and Discourses of Scientific Thought*, Moscow, Editorial URSS, 2000 (in Russian), p. 81. In another paper he writes: "The object of cosmology (in analogy with the objects of biology, cultural sciences and sociology) cannot be described within a single scientific discipline….From the point of view of the philosophy of science the universe represents an ideal object of theories pertaining to the human sciences, based in its construction in facts (astronomical observations and their interpretation) and related to the process of realisation of cosmologists' values and approaches, as well as to the discourse of the human sciences (for example the treatment of astronomical observations as characteristic texts and activity of the Cosmos)….(V. Rosin, "Towards a Problem of Demarcation of the Natural and Human Sciences, and where to One Must Relate Cosmology", *Epistemology and Philosophy of Science*, 2007, vol. XI, N 1, pp. 111-128 (In Russian).) Rosin makes his claims on the human-sciences' nature of cosmological knowledge by referring to works of another Russian philosopher V. Kazyutinski, in particular to his paper "Worlds of culture and world of science: an epistemological status of cosmology" in *A Socio-Cultural Context of Science*, Moscow: Institute of Philosophy of the Russian Academy of Science, 1998, pp. 101-118 (in Russian). However Kazyutinski himself objects to Rozin's strong claims on the status of cosmology as a human science. See his paper "No, cosmology is a physical science and not a human one", *Epistemology and Philosophy of Science*, 2007, vol. XI, N 2, pp. 125-129 (in Russian).

However, if one looks at the interplay from the point of view of their noematic poles, one must admit that the difference between them is always understood in terms of the radically distinct object domains outlined by the faculties of cognition. The natural sciences are characterised by the conviction that their subject matter is always "an object", and, in particular, a non-human object (whose principle of existence is not related to subjectivity and personhood), so that its reading does not require any mutual agreement or reciprocity apart from common substance based connotations (consubstantiality).[94] It is in this sense that if cosmology pretends to be consistently a natural science, it must fulfil the major requirement: the "object" of cosmology (allegedly the universe as a whole) must be "at distance" from subjects of knowledge and thus, in a way, to be inhuman, whose contingent existence manifests *itself from itself*, and not conditioned by the constituting human subjectivity that is devoid of the noetic carriers.[95] However, this demand creates tension with the fact that the very constitution of any object is performed by a particular operation of reason, which in spite of its imposing detachment from an object still remains behind it. Correspondingly the noetic pole in predications of the universe can be removed only in particular applications of cosmology (its astronomical part) desiring to deal with particular distinct physical objects which do  not have immediate impact on being of man, namely remote planets, stars or galaxies.  Such objects are characterised by persistent identity through a span of historical time and their appearance in human experience is not a construction, but an empirical fact. It is this identity which gives them the status of objectively existent entities. Contrary to these, some cosmological "objects" are simply constructions because they are observed as wholes only from this particular location and cannot be treated as objects independently of this fact: this applies first of all to clusters of galaxies which consist of "galaxies", which are at different distances from us and thus at different times (with respect to us), so that the question of the status of the cluster of galaxies as a distant and distinct *"object" with fixed spatio-temporal characteristics* (this is usually implied in physics and natural sciences) does not have sense – this "object" is a mental construction.[96] While introducing a construct of  a "cluster" of galaxies on the basis of the manifested phenomena,  a different intentionality is invoked which unifies different aspects of this "cluster" (different galaxies which do not exhibit directly any physical causation) in one "physical object" assigning to it such existence *as if* it is based on physical interaction (causation) of its parts. Here one can see that the language of intentionality (pertaining to the human sciences) cascades towards the language of physical causes (pertaining to the natural sciences).  This shows that the ideal of the natural sciences is not only problematic on the scale of the whole universe, but, in fact, on the scale of its "elementary  constituents" such as  clusters of galaxies.[97]

---

[92] C.f. C. W. Harvey, "Natural Science is Human Science. Human Science is Natural Science: Never the Twain Shall Meet". In B.E. Babich, et al (eds.) *Continental and Postmodern Perspectives in the Philosophy of Science*, p. 122.

[93] Cosmological research is driven by cultural and social factors, even by fashion (See, for example, R. Penrose, *The Road to Reality*, pp. 1017-20). Some authors claim that it represents a sort of ideology (M. Lopez-Corredoira, M., "Sociology of Modern Cosmology", arXiv:0812.0537v1 [physics.gen-ph] 2 Dec 2008.)

[94] E. Husserl accentuated a feature of "corporeity" which physics (as a typical representative of the natural sciences) only wants to see in that world from which this same physics originates, that is the life-world: "The natural science of the modern period, establishing itself as physics, has its roots in the consistent abstraction through which it *wants* to see, in the life-world only corporeity. Each "thing" "*has*" corporeity even though, if it is (say) a human being or a work of art, it is not merely bodily but is only "embodied", like everything real." (E. Husserl, *The Crisis of European Sciences and Transcendental Phenomenology*, p. 227).

[95] An attempt like this can be found in M. Tegmark's approach for whom the epistemic exhaustibility of the physical universe is equivalent to its description in a purely mathematical form devoid of "human baggage". (See his "Mathematical Universe").

What happens then is that the same shift from intentionality to causality takes place in creating the idea of the universe as a whole, when the appearance of filaments of clusters of galaxies through observations and conscious articulation is referred to the universe as a whole understood as a singular entity allegedly unified on the basis of physical causality. Here the language of intentions (in this case a *belief* in existence of the overall physical totality) is transformed into the language of physical causality. The intention is to unify (on the basis of a successive theoretical synthesis) the empirical images of causally disconnected regions in the sky in one single whole. But this unification naturally cannot be achieved as an accomplished phenomenalisation; for if the thus constructed unity is formulated, it cannot be phenomenalised simply because it does not belong to the same series of empirical appearances which were passed over by the theoretical synthesis. Such a synthesis distances itself from the contingently given appearances towards a simplified mathematical construction thus explaining away the problem of contingent givenness of it empirical references (a typical example is the cosmological principle which equates all positions in space thus making irrelevant the question on the contingency of the empirical display as it is given to us here and now: the universe is uniform and the same display would be in any position in space). Correspondingly, such a construct, represented by a global space-time diagram, being a logical digest of the variety of appearances reveals itself as poor in intuitive donation in the same sense as all mathematical constructions are: it effectively doe not produce any growth of knowledge apart from stating that we belong to the whole. Consequently the construct of the universe as a whole, while contributing to the constitution of the universe, does not explain the facticity of this particular constitution as contributing toward the facticity of the universe in general. Even less does it provide us with any insight on the physical causality in this whole. But it is this causality which is the object of desire of cosmologists. Unable to address this causality as being beyond phenomenalisation at present, cosmology makes its intentional objective that of producing a model where the unity of "all in all" in the universe would be explained in terms of *physical causality* but related to the past of the universe (where all causally disjoint regions were unified).[98] The move of thought is quite clear here: to assert the unity of the universe in terms of its absolute origin, that is to introduce physical causality among its presently contingent displays by referring them back to the point where "all was in all", that is physically connected simply through its belonging to "one and the same" (consubstantiality through origination). Thus the introduction of the idea of the evolving universe, being certainly supported by all known evidences from observational astronomy, and the shift to the past of the universe in order to interpret its contingent present, manifests an *epistemic causation* from "intentions" to "physical causes".

---

[96] That fact that intentionality plays here a pivotal role can be realised through an observation that a cluster of galaxies as a correlate of this intentionality remains unfulfilled on the level of physical causality. A cluster cannot be conceived as a physical system or object whose components are in physical interaction which are constitutive of this object. There is no "body" of the cluster of galaxies in the same way as there is a body of a train whose appearance through its front entails the assurance in its physical objectivity as a "solid body" localised in space and time.  One can conjecture  that the question of existence of such objects as clusters of galaxies is established through insistence on their epistemic identity and not space-time attributes.

[97] This resonates with Ellis's qualification that "cosmology is both a geographic and a historical science combined into one: we see distant sources at  an earlier epoch, when their properties may have been different."( G. Ellis,  "Issues in the Philosophy of Cosmology", p. 1221.)

---

[98] Characteristically the work of intentionality presupposes a sort of transcendence. Intentionality gazes beyond things' appearance, transcends the visible towards a non-visible, that is towards that which is not reducible to the visible and yet that which is the condition of it. Intentionality thus implies a speculative transcendence similar to that one of creative mythology inherent in the very human condition and may be having an evolutionary importance. The appeal to the past of the universe which "unites" the phenomenal in some unformed and undifferentiated matter means at the same time the invocation of the original time which in its actuality is infinitely far away but still active and present as an open-ended fulfilment. Being non-human this past is in radical discontinuity with this world, but it serves as a productive act in relation to the visible world, that is as the world at distance, that distance which can be crossed over but not immanently overcome. The transcendence in this case is a free flight from the possible to that which is the condition of any possible. At the same time transcendence is not arbitrary

Such a unification of all different aspects of the universe appearances in one "original" substance implies, however, not causal connection based on physical processes (contemporary cosmology is clear about the fact that the universe consists of space-time disconnected sections[99]); here one means connectedness as belonging to the underlying foundation (be it the overall encompassing space-time structure or substance), as consubstantiality of everything in the universe as a whole.[100] It is this type of consubstantiality that is implied when in some textbooks on cosmology the universe, containing according to relativity infinitely many causally disconnecting regions, is depicted in a single diagram meant to symbolise the totality of all.[101] Unlike a consubstantiality related to micro-particles constituting all physical objects, the large-scale cosmological consubstantiality does not have a clear image-like representation apart from mental diagrams. Thus this consubstantiality has rather a transcendental character referring to the conditions of knowledge of the universe as a whole. Correspondingly this consubstantiality is a product of intentionality which, in cosmology, cascades towards physical causality as cosmology desires to assign to this

---

fantasy and a lapse into the inarticulate, it is the intentional thematisation of that which makes the universe a unity, that is, in an ancient Greek parlance, the cosmos.

[99] This fact is related to the potentially infinite geometry of space and the finitude of the speed of light. Since we observe the universe along the past light cone which imposes constraints on the maximal distance causally connected with the earth bound point of observation, we are not only detached from all regions which are beyond this light cone, but event within this light cone we effectively receive signals from regions which, according to the standard model of cosmology, have been disconnected in the past. This constitutes a famous horizon problem, whose alleged solution was assigned to inflationary cosmology. (See e.g. S. Weinberg, *Cosmology*, pp. 205-6).

[100] One must not understand substance straightforwardly in the style of an old fashion metaphysics. When, for example we talk about the unity of the universe in the Big Bang, we assume a sort of unified field which contains potentially all differentiated objects. This assumption, for example, corresponds to an old Greek idea of "water" as that underlying agency which gives rise to all varied forms of matter.

[101] The examples of such diagram can be found in many standard books on cosmology; see, for example, E. R. Harrison, *Cosmology: The Science of the Universe*, Cambridge University Press, 2000, pp. 345-355.

consubstantiality an explicit physical meaning. It is not difficult to grasp that the transition from the language of intentions to the language of physical causality cannot be made on strictly scientific grounds for consubstantiality is not an empirical fact. It rather implies *faith in the existence* of the universe, or the world where both words carry connotations of the overall totality and unity. It is this faith that delivers us the sense of the *given* when we use the term "universe" in the conditions when the givenness pertaining to the natural object is unattainable. Using Husserl's words, "it is this *universal ground of belief in a world* which all praxis presupposes, not only the praxis of life but also the theoretical praxis of cognition. … *Consciousness of the world is consciousness in the mode of certainty of belief*; it is not acquired by a specific act which breaks into the continuity of life as an act which posits being or grasps the existent or even as an act of judgement which predicates existence. All of these acts already presuppose consciousness of the world in the certainty of belief."[102] These beliefs correspond to that which in the natural attitude can be described as empty and never fulfilled intentions. Then the very tendency to transform the language of intentions into the language of physical causes in the context of the universe as a whole represents an attempt to make the universe a target of ever-going but unfulfilled intentions.[103]

On the one hand the notion of the universe comes from astronomical observations and theories based in the ideal of physical causality; on the other hand we have some stories of the Big Bang and the universe's facticity expressed in philosophical enquiries and scientific-mythological narratives guided by the language of intentions and having origination in the human condition. This all suggests that either region in cosmological discourse (its observational base as well as eidetic extrapolations) encompasses the other; in other words, each type of cosmological understanding accounts for the hidden unity of both intentionalities dwelling in one and the same human person who discloses the universe. Put differently, cosmological theories need inputs from existential faith and hence from philosophy (regulative ideals in a Kantian sense), whereas philosophical imagination in the creation narrative borrows and exploits, for its "visualisation", physico-mathematical images thus offering a metaphysical extension of physics. [104]

---

[102] E. Husserl, *Experience and Judgement*, London: Routledge & Kegan Paul, 1973, p. 30. One must be aware, however, that Husserl's usage of the term "world" does not correspond exactly to what is meant here by the universe. He does not reduce the meaning of the world to the all-encompassing extended spatiality and temporality, but rather means the world as an irreducible context of all experience, as the "horizon of all horizons" in all intentional acts. (See in this respect a classical paper by L. Landgrebe, "The World as a Phenomenological Problem", *Philosophy and Phenomenological Research*, Vol. 1, 1958, pp. 38-58.)
[103] Cf. R. Sokolowski, *Introduction to Phenomenology*, 2000, p. 43.

One now anticipates that any attempt of totalising the world view, that is making a unique and consistent whole in our perception of the universe, is doomed to fail. All these attempts start from within the life-world associated with a geocentric world, and it is the life world which remains patchy and incoherent through different articulations including not only philosophical and scientific, but also religious ones. In other words, the life world does not allow its totalisation either through the language of matter and body or that of spirit and soul, through physical causes or through human intentions because it is the world of historically contingent *events* whose instantiation is not subject to the physical or purely spiritual.[105] The reality of the life world is far too existentially complex to allow a simple-minded reduction to either one of these. This implies that cosmology has to deal with this intrinsic dualism between its orientation towards the natural sciences and, at the same time, its dependence upon the dimensions of human life. One must then expect that the discursive language of physical causation as a mode of thematisation of the life-world will go side by side with the "language of *communion*" and the excess of intuition over reason, so that neither of them will be able to reduced to the other. Every attempt to semantically transgress the normalised sense-borders of everyday intentional life in the conditions of inescapable communion with the universe, by the use of causally reductive language which sees that universe in stages of evolution and hierarchy of objects, issues in a counter-sense. The same holds for attempts to transgress the causal domain with intentional language: intentional language is useful to analyse and refer causal language to existential motivations of research, but not to the truth of a *fact*, which causal language attempts to *affirm*.

---

[104] As an example of such a metaphysical extension of cosmology one can point to diagrams which recapitulate the wholeness of the universe (with some particular physical details) and its link to the fact that it is articulated by human beings. As an example one can point to a famous "closed circuit" in J. A. Wheeler's writings symbolising the world as a self-synthesising system of existence (See e.g. "World as a System Self-Synthesized by Quantum Networking." *IBM Journal of Research and Development*, 32, 1988, pp. 4–15 (5)), or a picture of the so called "Cosmic Uroboros" (See e.g. J. Primack, N. E. Abrams, *The View from the Centre of the Universe*, pp. 160, 284). These diagrams mean to stand for such a unity of the world in which the historically formulated physics is erected to the level of apodictic structure of being. One then understands the apophatic sense of these representations providing signifiers of the universe's manifestation with no pretence for the exhaustion of the sense of that which is signified.

[105] As was argued by J. L. Marion by referring to Kant, unique occurring typical for historical phenomena do not fall under the rubric of *analogies of experience* which concern only a "fringe of phenomenality typical of the objects constituted by the sciences, a phenomenality that is poor in intuition. (See J.-L. Marion "The Saturated Phenomenon", in D. Janicaud et al., *Phenomenology and "The Theological Turn": the French Debate*. New York: Fordham University Press, 2001, pp. 176-216 (204)).

As an example of how the language of intentionality takes control over the language of causality one can refer to the analogy between the phenomenology of birth (as absolute coming into being of a new hypostatic existence) and the phenomenology of the Big Bang: in both cases the "event" of birth and the origin of the universe are phenomenologically concealed because of the immanence of human life to itself as well as to the universe: one cannot transcend one's own life or the universe in order to "look" at their origin from outside. However this analogy, originating essentially at the level of intentionality, does not cascade towards the explanatory level of physical causation. It just points towards the fundamental limit in attempting to assign physical causation to the event of origination of the universe which originates from the fact of human embodiment in the universe.[106]

This analogy between the phenomenology of birth and the origin of the universe elucidates, in a non-trivial way, the sense of communion with the universe.[107] By referring to the mystery of origin of every particular personal existence, a physical problem of comprehending the temporal origin of the universe is transformed into a purely philosophical problem of the contingency of this origin, that is the problem of the sufficient ground for the whole temporal span of the universe. Using the analogy with famous Kantian antinomies one can say that the classical paradox of the temporal origin of the universe formulated by Kant in his first cosmological antinomy is shifted toward the antinomy of the absolutely necessary being as if it is the ground of the visible universe.[108] This transformation, which is not a result of the advance of the

---

[106] See more details in A. Nesteruk, *The Universe as Communion*, pp. 247-50.

[107] The sense of communion implied here exceeds its physical dimension of consubstantiality. The analogy which is developed by us differs from the hypothesis of "genetic similarity" between the evolution of the universe (cosmogenesis) and development of a human being (anthropogenesis) introduced by a Russian philosopher A. Pavlenko. (See for example, A. N. Pavlenko, "A Place of 'Chaos' in the New World 'Order'", pp. 47-48; "Universalism and Cosmic Harmony: A Principle of Genetic Similarity", pp. 389-401; "European Cosmology: between 'birth' and 'creation'", in A. A. Grib (ed.) *Scientific and Theological Thinking of Ultimate Questions: Cosmology, Creation, Eschatology*, Moscow, St. Andrew Biblical Theological Institute, 2008, pp. 128-130 (in Russian)).

[108] This shift can be observed by analysing R. Penrose's hypothesis on the origin of temporal irreversibility due to the local conditions and treated by him through a pseudo-theological metaphor (See his *The Emperor's New Mind*. New York: Oxford University Press, 1989, pp. 435-447.) See details in A.

physical sciences but the work of the intentionality enquiring into the ground of its own facticity and hence the facticity of the universe, demonstrates that from within this intentionality the problem of the temporal origin of the universe and its explication through evolutionary stages is irrelevant for any attempt to understand the universe's facticity.[109] The analogy with the phenomenological concealment of the origin of personal consciousness is crucial here. When achieving such a state of consciousness, when all the historical, temporal and spatial contingent aspects of the universe are reduced, this consciousness has to embrace itself to the uncertain infinity of its own being. Being absolute as an event of existence, every personal consciousness treats itself as indefinite and commensurable with the universe, where the commensurability manifests itself as an intuition of co-existence with the universe, which is not fulfilled through acts of reason in spatial and temporal distinctions. According to G. Marcel's thought expressed at the beginning of the 20[th] century, "the universe as such, not being thought of or able to be thought of as an object, has strictly speaking no past: it entirely transcends what I called a 'cinematographic' representation. And the same is true of myself: on a certain level I cannot fail to appear to myself as contemporary with the universe (*coaevus universo*), that is, as eternal."[110] Marcel anticipates here a simple truth that for every human being the sense of communion with the universe makes existentially irrelevant any notion of evolution and stage-by-stage description of the universe ('cinematographic' representation).[111] Communion is here and now and it is

---

Nesteruk, "Temporal Irreversibility: Three Modern Views." In *Time, Creation and World-Order*. Ed. M. Wegener, Aarhus University Press, 1999, pp. 62–86 (77-78); *Light from the East: Theology, Science and the Eastern Orthodox Tradition*. Minneapolis: Fortress Press, 2003, pp.67-77.

[109] If one thinks of the ontological "transcendental condition" for the possibility of consistent and varied processes in time in the universe, then no appeal can plausibly be made either to a set of fixed entities or to an ontic first cause since, within the material universe, neither the former nor the latter can "originally", that is ontologically, precede what is caused. The notion of 'cause' is acting here as a pragmatic fiction which disguises the fact that 'that which causes' is only something which is changing into something else. In this sense all ultimate causes are more primarily effects, that is an effect is described as *causatum* (rather than as *effectus*). In different words one expresses the same by saying that "causing means giving" implying the cause is a going out of itself as an effect, while the effect is wholly 'from' the cause in which it eminently abides. The sought explanation of the facticity of the universe thus implies the imminence of "givenness", that is that donation of the universe which we receive. Correspondingly all conventional mythologies acting on the premise of imitating the other-worldly "origin" of the universe as a precondition for the unfolding temporal flux fail to address this contingency in an "explanatory" manner: they just imitate the mode of donation of the universe through the appeal to creative power of imagination which allegedly refers to realities beyond the facticity of the donation. In other words, the donation itself is explicated in terms of layers and strata of this same donation which is not perceived as manifested.

[110] G. Marcel, *Being and Having*, p. 24.

[111] The so called "cinematographic representation of the universe" (the term which goes back to H. Bergson) is irrelevant in the approach of communion even in spite of a trivial intuition that all material constituents of communicants contain elements of the historical past of the universe. Through its past the

absolute in its transcendent phenomenality as an event of life; in other words it is the facticity of life that retains transcendence of that with which and in which this life is as communion. Life implies internal time consciousness, its unending inevitability, but being integrated in one and the same person this consciousness is not obliged to be projected on the physical extension of time.    If the universe in its phenomenality appears to be a stable and  enduring  background of existence then a human being as a communicant with the universe realises itself as commensurable with it and hence also co-eternal with it. One can say that  human beings as long as they are alive experience the immanence of the infinite. The life of a human being is then an act of communion, an event whose fullness (perceived through the sense of living being) does not need any acknowledgement of history of the universe, although the impossibility of grasping the sense of facticity of this life perceived through the immanence remains the unavoidable *negative certitude* of a theological kind.  Within the fact of life  it is the universe that becomes a part of existential history of every human being and not vice versa.[112] Thus the identity of the universe which receives its fulfilment in acts of communion with persons, transcends any phenomenalisation of the universe (as its representation) through cosmology.[113]  However one must admit that the irrelevance of the non-lived cosmic history for a particular event of life does not entail the irrelevance of the *human history* as related to lived moments and memories imbued with a sense of telos related to the infinite tasks of humanity.

What happens in cosmology then is that physicists' intentionality (that is that one which pertained to the natural attitude) breaks the noetico-noematic (subject-

---

active communion with the universe manifests itself as being present  in us in every cell and every breath.

[112] This observation corresponds to a phenomenological stance on physical time as originating in the subjective time as well as in the internal time consciousness. (See E. Husserl, *On the Phenomenology of the Consciousness of Internal Time* (Tr. by J. B.  Brough), Dordrecht, Boston, London, Kluwer Academic Publishers, 1991).  It also connotes with a problem of constitution of the sense of history in transcendental ego. (See in this respect Ricoeur, P. *Husserl. An Analysis of His Phenomenology*. Evanston, Northwestern University Press, 1967, pp. 145-150).

[113] The epistemological dichotomy between communion and discursive representation of the universe implied here can be expressed as the opposition between *subjective absolute* and *objective relative* mentioned by H. Weyl in reference to M. Born: "The immediate experience [as communion, A.N.] is *subjective and absolute*. …The objective world, on the other hand with which we reckon continually in our daily lives and which the natural sciences attempt to crystallise…is of necessity *relative* [this relativity is first of all related to its historical contingency, A.N.];  it can be represented by definite things (numbers or other symbols) only after a system of coordinates has been arbitrarily [that is historically contingently, A. N.] carried into the world….Whoever desires the absolute must take the subjectivity and egocentricity into the bargain; whoever feels drawn toward the objective faces the problem of relativity" (H. Weyl,  *Philosophy of Mathematics and Natural Science*,  p. 116.)

object) inseparability and explicates the event of communion with the universe through creating theoretical models of the universe. By being in communion the knowing subject, a cosmologist, articulates the universe as a sort of "out there" which allegedly follows the objective laws of physics and thus is independent of a cosmologist's insight. Such a scientific notion of the universe naturally falls under the phenomenological critique which reinstates back a simple truth that any truth of the universe is an articulated truth, so that this truth is in man and his body, the body which is consubstantial to the universe and communes with it. Thus one comes back to the inevitability of the link between the universe and man in a deep philosophical sense. However, and this is important for us, this intertwining between man and the universe does not deprive the universe of independence from the conditions of its expression by the human subjectivity. The universe as communion, denying any complete synthesis in its phenomenalisation, always retains that overwhelming presence which cannot be conditioned by the rubrics of subjectivity, inducing an excess of intuition over any attempt to see or constitute them. Thus the universe appeals to man as a saturated phenomenon always retaining its own transcendence with respect to all attempts of humanity to grasp the sense of its facticity.[114]

In conclusion one has to confirm the main thought of this paper that there is an obvious and probably unavoidable tension between the representation of the universe as an object in cosmology and its presence in existential communion which affects all our attempts to express the experience of living in the universe. To avoid this tension, one has to step into a dispassionate *phenomenological* description of the universe as it discloses itself in the natural attitude, in the language of causes on the one hand and in the language of intentional immanence through communion on the other hand. One might say that the *ontological commitment* should be left out and a certain phenomenological calm must be adopted with respect to various languages used for assessing the universe as a whole. This means that one can use discoveries achieved by natural scientific reduction as well as by philosophical insight and communion without committing semantically or ontologically to one region's priority over the other. The same idea can be expressed differently: the cosmological narrative follows either the logic of physical causes and is shaped by mathematics aiming at pure objectivity or, alternatively, the "logic" of life and inseparable communion with the universe (which is not subject to intellectual persuasion and thus is the free-willing employment of artistic expression), which is ever incomplete (metaphorical) and fundamentally open-ended. It is because of this dichotomy that one must learn how to live with incomplete wholes, partial and shattered totalities – totalities requiring different languages although, after all, belonging to one culture.[115] Correspondingly the objective of a philosophical insight in cosmology, undertaken here, is not to find a unified language for understanding the universe, but rather to realise that in our approach to its totality, always initiated in the life-world, we progress by the various ways given to humanity. The reality of the universe then is much more than is met by the discursive mind, it forms a mysterious sense of "identity", which is intuited, but never completely grasped by the mind: it bedazzles us, while constituting our own sense of identity to the extent that we cannot circumscribe the universe in the rubrics of thought. In its perennial leap towards understanding the sense of the universe, humanity stretches its capacity to grasp itself. Indeed, ontologically, the universe we disclose through our embodiment involves us as disclosing it. Given this our personhood achieves its status as a "place" that permits disclosure of the universe through what this personhood is. The *I* as person discloses, by being structured by disclosure itself, that is, in the words of M. Merleau-Ponty, by being a "concrete emblem of general manner of being."[116]

---

[114] The notion of the "saturated phenomenon" was introduced by J. L. Marion in his paper "The Saturated Phenomenon", and developed further in his books *Being Given*, and *In Excess*. The application of the idea of saturated phenomena in cosmology is undertaken by A. Nesteruk in his paper "Transcendence-in-Immanence in Theology and Cosmology, 186-189.

---

[115] C.f. C. Harvey, "Natural Science is Human Science", p. 133. See also L. Papin, "This is not a Universe: Metaphor, Language, and Representation", *Proceedings Modern Language Association*, vol. 107, No. 5, 1992, pp, 1253-65. The implied diversity and plurality, as a valid approach to knowledge of the universe, brings to a characteristic expression that, in cosmology, all signifiers in our experience of the universe do not exhaust that which is signified. As was conjectured by D. Bohm and D. Peat, each scientific theory bears the inscription "this is not a universe" meaning that "every kind of thought, mathematics included, is an abstraction which does not and cannot cover the whole of reality" and this is why "perhaps every theory of the universe should have in it the fundamental statement 'this is not a universe'" (D. Bohm, D. Peat, *Science, Order, and Creativity*, New York, Bantam, 1987, pp. 8-9).
[116] M. Merleau-Ponty, *The Visible and the Invisible*, Evanston, Northwestern University Press, 1968, p. 147.

## Acknowledgements

I would like to express my feelings of gratitude to George Horton, Christopher Dewdney, David Matravers, Mogens Wegener, Joel Mathews, Grigory Goutnerr and Ruslan Loshakov for discussion and helpful comments.